\documentclass[fleqn,usenatbib]{mnras}

\usepackage{newtxtext,newtxmath}
\usepackage[T1]{fontenc}
\usepackage{ae,aecompl}

\usepackage[utf8]{inputenc}
\usepackage{amsmath}
\usepackage{graphicx}
\usepackage{amsfonts}
\usepackage{xcolor}
\usepackage{verbatim} 
\usepackage{caption}
\usepackage{subcaption}
\usepackage{ulem}
\usepackage{anyfontsize}

\newcommand{\orcid}[1]{\href{https://orcid.org/#1}{\includegraphics[scale=0.08]{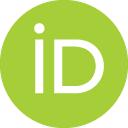}}}

\newcommand{\mvir}{M_{\rm 200c}}

\newcommand{\rvir}{R_{\rm 200c}}

\newcommand{\mgassat}{M_{\rm gas}^{\rm sat}}
\newcommand{\mcgas}{M_{\rm CoolGas}}

\newcommand{\mcgasicm}{\mcgas^{\rm ICM}}

\newcommand{\mstar}{M_\star}
\newcommand{\mstarsat}{\mstar^{\rm sat}}
\newcommand{\rhalfstar}{R_{\rm half,\star}}

\newcommand{\sublink}{\textsc{sublink} }
\newcommand{\sublinkgal}{\sublink\_\textsc{gal} }
\newcommand{\subfind}{\textsc{subfind} }

\newcommand{\tvir}{T_{\rm vir}}
\newcommand{\tcool}{t_{\rm cool}}
\newcommand{\tff}{t_{\rm ff}}

\newcommand{\msun}{{\rm M}_\odot}

\DeclareRobustCommand{\ion}[2]{%
\relax\ifmmode
\ifx\testbx\f@series
{\mathbf{#1\,\mathsc{#2}}}\else
{\mathrm{#1\,\mathsc{#2}}}\fi
\else\textup{#1\,{\mdseries\textsc{#2}}}%
\fi}

\graphicspath{{figures/}{figures/CoolGasMaps}{figures/SFRMaps}}
\DeclareGraphicsExtensions{.pdf,.png,.svg,.gif}
\DeclareCaptionFormat{cont}{#1 (cont.)#2#3\par}

\title[The cooler past of the intracluster medium]{The cooler past of the intracluster medium in TNG-Cluster}
\author[Rohr et al.]{Eric Rohr$^{1}$\thanks{Contact e-mail: \href{mailto:rohr@mpia.de}{rohr@mpia.de}}\orcid{0000-0002-9183-5593},
Annalisa Pillepich$^{1}$\orcid{0000-0003-1065-9274}, Dylan Nelson$^2$\orcid{0000-0001-8421-5890}, 
Mohammadreza Ayromlou$^{2,3}$\orcid{0000-0003-3783-2321},
\newauthor
C{\'e}line P{\'e}roux$^{4,5}$\orcid{0000-0002-4288-599X},
Elad Zinger$^{6,1}$\orcid{0000-0002-6316-3996}
\\
\\
$^{1}$Max-Planck-Institut f{\"u}r Astronomie, K{\"o}nigstuhl 17, D-69117 Heidelberg, Germany\\
$^{2}$Zentrum f{\"u}r Astronomie der Universit{\"a}t Heidelberg, ITA, Albert Ueberle Str. 2, D-69120 Heidelberg, Germany\\
$^{3}$Argelander-Institut f\"ur Astronomie, Auf dem H\"ugel 71, D-53121 Bonn, Germany\\
$^{4}$European Southern Observatory, Karl-Schwarzschildstrasse 2, D-85748 Garching bei M{\"u}nchen, Germany\\
$^{5}$Aix Marseille Universit{\'e}, CNRS, LAM (Laboratoire d’Astrophysique de Marseille) UMR 7326, F-13388 Marseille, France\\
$^{6}$Centre for Astrophysics and Planetary Science, Racah Institute of Physics, The Hebrew University, Jerusalem 91904, Israel \\
}

\date{}

\pubyear{2024}

\begin{document}
\label{firstpage}
\pagerange{\pageref{firstpage}--\pageref{lastpage}}
\maketitle

\begin{abstract}
The intracluster medium (ICM) today is comprised largely of hot gas with clouds of cooler gas of unknown origin and lifespan. We analyze the evolution of cool gas (temperatures~$\lesssim10^{4.5}$~K) in the ICM of 352 galaxy clusters from the TNG-Cluster simulations, with present-day mass $\sim10^{14.3-15.4}\,\msun$. We follow the main progenitors of these clusters over the past $\sim13$~billion years (since~$z\lesssim7$) and find that, according to TNG-Cluster, the cool ICM mass increases with redshift at fixed cluster mass, implying that this cooler past of the ICM is due to more than just halo growth. The cool cluster gas at $z\lesssim0.5$ is mostly located in and around satellite galaxies, while at $z\gtrsim2$ cool gas can also accrete via filaments from the intergalactic medium. Lower-mass and higher-redshift clusters are more susceptible to cooling. The cool ICM mass correlates with the number of gaseous satellites and inversely with the central supermassive black hole (SMBH) mass. The average number of gaseous satellites decreases since $z=2$, correlating with the decline in the cool ICM mass over cosmic time, suggesting a link between the two. Concurrently, kinetic SMBH feedback shifts the ICM temperature distribution, decreasing the cool ICM mass inside-out. At $z\approx0.5$, the predicted \ion{Mg}{ii} column densities are in the ballpark of recent observations, where satellites and other halos contribute significantly to the total \ion{Mg}{ii} column density. Suggestively, a non-negligible amount of the ICM cool gas forms stars in-situ at early times, reaching $\sim10^{2}\,\msun\,{\rm~yr^{-1}}$ and an H$\alpha$ surface brightness of $\sim10^{-17}\,{\rm~erg\,s^{-1}\,cm^{-2}\,arcsec^{-2}}$ at $z\approx2$, detectable with Euclid and JWST. 
\end{abstract}



\section{Introduction} \label{sec:intro}

The nature of gas in and around cluster of galaxies, the intracluster medium (ICM), is sensitive to both the cosmology and feedback processes of the cluster members, providing one of the best astrophysical laboratories for testing our theories of hierarchical structure formation and galaxy evolution. The ICM today is mostly hot at temperatures $\sim10^{7.5-8}$~K that emit via thermal bremsstrahlung emission in the X-ray \citep[e.g.,][]{Sarazin1986,Bulbul2024}. The ICM is, however, not a single temperature fluid but instead multiphase in nature, both from a theoretical and observational perspective \citep{McCarthy2004,Olivares2019}. Beyond the hot X-ray emitting plasma, local clusters have been shown to contain multiphase cloudlets of warm-hot $\sim10^{4.5-6}$~K, cold-cool $\sim10^{4-4.5}$~K gas, and/or molecular $\sim10^{2}$~K gas, which may be observable via thermal and kinetic Sunyaev-Zel'dovich effect \citep{Mroczkowski2019}, H$\alpha$ filaments and nebulae \citep{Fabian2003,Crawford2005}, \ion{H}{i} and \ion{Mg}{ii} emission and absorption in background quasar spectra \citep{McNamara1990,Lanzetta1995,Chen1997}, and molecular CO emission in both the central galaxies and in cooling filaments \citep{Salome2006,McNamara2014,Omoruyi2024}. At higher redshifts $z\gtrsim2$, clusters and their progenitors, or protoclusters, have already assembled their hot ICM \citep{Tozzi2022,DiMascolo2023}, which still contain multiphase gas observable as Ly$\alpha$ nebulae \citep{Steidel2000,Matsuda2012}, optical absorption features \citep{Prochaska2013}, and even radio emission from cold, molecular gas \citep{Chen2024}. In fact, some observations of gas cooling in local and higher redshift $z\gtrsim 1$ clusters suggest that the ICM may even be able to cool to the point of forming stars \citep{McNamara1989,Webb2015,Hlavacek-Larrondo2020,Barfety2022}. At even higher redshifts $z\gtrsim 4$, there may exist large reservoirs of neutral gas in the proto-cluster ICM, as inferred by Ly$\alpha$ absorption \citep{Heintz2024}. The exact nature of cool halo gas and how it evolves with cosmic time remains largely unknown. 

Cluster progenitors form in the early Universe at the strongest density perturbations in the centers of their dark matter halos. They grow by accreting material -- namely dark matter, gas, and smaller satellite galaxies or subhalos -- from the intergalactic medium \citep[IGM; e.g.,][]{Springel2001,Springel2005b}. Infalling gas is thought to be shock heated to the virial temperature of the halo, forming a hydrostatic halo supported by thermal pressure \citep{Rees1977,Silk1977,White1978}. However, halos less massive than some critical threshold mass $\sim10^{12}\, \msun$ at $z\sim2$ cannot sustain hot atmospheres \citep{Binney1977, Birnboim2003,Katz1991,Fardal2000}. Simulations suggest that cool- or cold-mode accretion is likely the main source of growth for lower-mass halos, followed by a hot-mode accretion for higher-mass halos \citep[e.g.,][]{Keres2005,Dekel2006,Keres2009,Faucher-Giguere2011,Nelson2013}, although the exact details depend on numerics and baryonic feedback (\citealt{VandeVoort2011,Nelson2015,Mitchell2020}; see also \citealt{Vogelsberger2020,Crain2023} for recent reviews of galaxy formation models in cosmological simulations). Note however that these works focus on lower halo masses at the group mass scale and below, whose evolution may not be directly comparable to cluster progenitors.

There are a number of additional physical complications that impact the multiphase nature of the ICM. Hot halo gas can cool down due to thermal instabilities enhanced by local density perturbations \citep[e.g.,][]{sharma2012a,McCourt2012,Voit2017,Choudhury2019}. For example, using the cosmological magneto-hydrodynamical (MHD) simulations from the IllustrisTNG project, \citet{Nelson2020} and \citet{Ramesh2023b} show that the passage of satellites provides strong density perturbations, triggering gas cooling in the halo gas of luminous red galaxies at intermediate redshifts and in the circumgalactic medium (CGM) of Milky Way-like galaxies today. Furthermore, gaseous satellites may deposit their mostly cold interstellar medium (ISM) directly into halos, which has been ubiquitously observed in jellyfish galaxies \citep[e.g.,][]{Cortese2006,Poggianti2017,Roberts2021b,Cortese2021,Boselli2022} and recently suggested in cosmological simulations \citep{Rodriguez2022,Rohr2023,Weng2024,Chaturvedi2024}. We again note, however, that these numerical works have largely studied low-mass clusters and groups.

Feedback from the central supermassive black hole (SMBH) may also affect the halo baryons by redistributing gas, driving material from the central galaxy into the halo and beyond, heating the halo gas to super-virial temperatures, inducing turbulence as a form of non-thermal pressure support, and providing the perturbations to trigger gas cooling \citep[e.g.,][]{Li2014,Qiu2019,Beckmann2019,Nelson2019b,Zinger2020}. Recently, the Manhattan project, a suite of $\sim100$ zoom simulations of clusters with $z=2$ mass $\gtrsim10^{14}\, \msun$ predict large amounts of cool, neutral gas in the proto-clusters at high redshifts $z\gtrsim2-5.5$ \citep{Rennehan2024,Heintz2024}. Idealized MHD simulations of galaxy clusters have also shown that both SMBH feedback and magnetic fields may be important for the creation and survival of cool gas in clusters \citep{Fournier2024,Hong2024}. For the latter, there is an extensive body of literature studying a cool gas cloud in motion with respect to the ambient hot medium, that is, the cloud-crushing problem (e.g., \citealt{Li2020,Sparre2020,Fielding2022,Gronke2022}; this is typically studied from the perspective of a slow-moving cold gas cloud being accelerated by a fast wind, but the same physics applies to a cold gas cloud moving through the ICM). 

In this work, we analyze the history and evolution of the ICM, concentrating on the cool gas, using the TNG-Cluster project \citep[][\textcolor{blue}{Pillepich et al. in prep.}]{Nelson2024}. With its sample of 352 simulated high-mass galaxy clusters, we study the cool ICM of temperatures $\sim10^{4-4.5}$~K (and star-forming gas), where the IllustrisTNG galaxy formation model (TNG hereafter) has already been shown to naturally produce multiphase gas in the CGM and in group-mass halos \citep[][albeit at better resolution than in TNG-Cluster]{Nelson2020,Nelson2021,Ramesh2023a}.
We then focus on how the total cool ICM mass and the significance of the relevant physical processes evolve across cosmic time, and on how this is connected to both internal and global properties of the clusters. Our work combines a statistical study of the cool ICM in cosmological simulations of massive galaxy clusters from the first billion years until today including magnetic fields, hierarchical structure formation and infalling satellites, and SMBH feedback from a well-tested galaxy formation model.

The remainder of the paper is organized as follows. In \S~\ref{sec:meth} we describe the main methods, including the TNG-Cluster simulation, TNG galaxy formation model, and the cool ICM definition. In the results in \S~\ref{sec:results}, we present the evolution of the cool ICM over the past $\approx13$~billion years, since $z\sim7$. We then delve into the sinks and sources affecting the evolution of the cool ICM in \S~\ref{sec:why}. In \S~\ref{sec:disc} we discuss the observational implications of the cool ICM. Lastly in \S~\ref{sec:sum}, we summarize the main results and conclusions.


\section{Methods} \label{sec:meth}

\subsection{TNG-Cluster} \label{sec:meth_tng}

TNG-Cluster\footnote{\url{www.tng-project.org/cluster/}} is a suite of 352 massive galaxy cluster zoom simulations, spanning halo masses $\mvir\approx10^{14.3-15.4}\,\msun$ \citep[][\textcolor{blue}{Pillepich et al. in prep.}]{Nelson2024}. The technical details of TNG-Cluster are given in \citet{Nelson2024} and briefly summarized here. 

The re-simulated clusters were drawn from a $\approx1$~Gpc box-size parent dark matter only simulation based only on $z=0$ halo mass such that: (i) all $\approx 90$ halos with mass\footnote{In this work, we refer to the halo mass as $\mvir$, the mass enclosed by the halo radius $\rvir$ such that the total average enclose density is equal to 200 times the critical density of the universe at that time.} $>10^{15}\,\msun$ are included; and (ii) halos with mass $10^{14.3-15.0}\,\msun$ were randomly selected such that the halo mass distribution is flat over this mass range. 

The TNG-Cluster simulation employs the well-tested TNG galaxy formation model \citep{Weinberger2017,Pillepich2018b}. In short, the TNG and TNG-Cluster simulations evolve stars, SMBHs, gas, cold dark matter, and magnetic fields; the latter start as a uniform field of strength $\approx10^{-14}$~G and are allowed to grow and evolve thanks to dynamo effects, shear motions and turbulence \citep[e.g.,][]{Pakmor2011,Pakmor2013}, from $z=127$ until today. The simulations use the \textsc{arepo} code \citep{Springel2010,Weinberger2020}, where the fluid dynamics are discretized and solved on a moving Voronoi mesh. 

Gas heats and cools, including metal-line cooling, based on redshift, temperature, density, and metallicity following \citet{Wiersma2009} in the presence of a redshift-dependent UV background \citep{Katz1992,Faucher-Giguere2009}, and, for gas cells near SMBHs, the radiation field from the AGN \citep{Vogelsberger2013}. The relationship between density and temperature for dense $n > 0.1\, {\rm cm}^{-3}$ gas is determined via an effective equation of state \citep{Springel2003}, which assumes a two-phase ISM, where the cold phase dominates the mass and the hot phase dominates the volume, and the relative pressure contributions of the two phases are regulated by the unresolved stellar feedback. For this analysis we adopt $10^{3}$~K as the temperature of star-forming gas, cooler than the temperature floor of $\approx10^4$~K. Cool/cold gas denser than $n \gtrsim 0.1\, {\rm cm^{-3}}$ is able to form stars probabilistically assuming a Chabrier initial mass function \citep{Chabrier2003} based on the Kennicutt-Schmidt Law \citep{Schmidt1959,Kennicutt1998}. Stellar particles represent populations of stars, which return mass and metals to the neighboring gas via asymptotic giant branch stars, Type Ia, and Type II supernovae. Galactic winds, powered by Type II supernovae, eject star-forming gas in a hydrodynamically decoupled kinetic wind scheme \citep{Pillepich2018b}.

SMBHs are seeded at a mass of $\approx 1.2\times10^{6}\, \msun$ in the centers of halos with a total Friends-of-Friends mass $\gtrsim 7.5\times10^{10}\, \msun$, if they do not already have a SMBH. SMBHs are re-positioned to the minimum of the local gravitational potential, and SMBHs merge if they are within their neighbor regions \citep{Weinberger2017}. SMBHs grow both via such mergers and by accreting gas, via Bondi-Hoyle-Lyttelton accretion \citep{Hoyle1939,Bondi1944,Bondi1952}, where the maximum accretion rate $\dot{M}_{\rm SMBH}$ is set by the Eddington limit $\dot{M}_{\rm Edd}$. The feedback from SMBHs is bimodal, where the mode is determined by the Eddington ratio $f_{\rm Edd}$ and SMBH mass $M_{\rm SMBH}$. Specifically, the SMBH feedback is in the thermal, high-accretion mode when 
\begin{equation}
    f_{\rm Edd} \equiv \dfrac{\dot{M}_{\rm SMBH}}{\dot{M}_{\rm Edd}} \geq \chi,\ \chi = {\rm MIN}\left[0.002\left(\dfrac{M_{\rm SMBH}}{10^8\, \msun}\right)^2, 0.1\right].
\end{equation}
The SMBH is in the kinetic, low-accretion mode when $f_{\rm Edd} < \chi$. In the high-accretion mode, the feedback is via a continuous thermal energy injection to the local environment; in the low-accretion mode, the feedback is via discrete kinetic energy injections in random directions that change for each injection, such that the energy and momentum are conserved in a time-averaged sense \citep[][]{Weinberger2017}. As we will show below, low-redshift galaxy clusters in the TNG model are always in low-accretion mode and exert kinetic feedback.

The baryon mass resolution of TNG-Cluster is $m_{\rm bar} = 1.1\times10^{7}\,\msun$, the same resolution as TNG300 from the original TNG simulation suite \citep{Pillepich2018,Nelson2018,Naiman2018,Marinacci2018,Springel2018}. We note that the TNG galaxy formation model at TNG-Cluster mass resolution has already been at least partially validated in the low-mass cluster regime \citep[e.g.,][]{Nelson2018,Donnari2021a,Truong2020,Donnari2021b} and in the TNG-Cluster first results papers \citep{Ayromlou2024,Lee2024,Lehle2024,Nelson2024,Rohr2024,Truong2024}. In this analysis we only consider high resolution particles and cells, whose masses are similar to the target mass listed above and are located in the targeted zoom regions, that is, in or near the target clusters.

We adopt the same cosmology as TNG and TNG-Cluster, consistent with the \citet{Planck2016} results: $\Omega_{\Lambda,0} = 0.6911, \Omega_{\rm m,0} = \Omega_{\rm bar,0} + \Omega_{\rm dm,0} = 0.3089, \Omega_{\rm bar,0} = 0.0486, \sigma_8 = 0.8159, n_s = 0.9667, {\rm and}\ h = H_{\rm 0} / (100\, {\rm km\, s^{-1}\, Mpc^{-1}}) = 0.6774$, where $H_0$ is the Hubble parameter. 

\subsection{Cluster sample and ICM definitions} \label{sec:meth_sample}

In this work, we exclusively focus on the 352 primary zoom targets from the TNG-Cluster project. We refer to the central galaxy within each of these clusters as the brightest cluster galaxy (BCG). The galaxy stellar mass $\mstar$ is the stellar mass enclosed within twice the stellar half mass radius. We define the most massive supermassive black hole within each BCG as the main SMBH, which usually becomes the most massive SMBH in the BCG at $z=0$.

Dark matter halos are identified using the Friends-of-Friends (FoF) algorithm with a linking length $b=0.2$, run only using the dark matter particles \citep{Davis1985}. Then the baryonic components are connected to the same halos as their closest dark matter particle. Throughout this paper, we use ``FoF group,'' ``halo,'' and ``cluster" synonymously. Galaxies are then identified using the \subfind algorithm, which identifies gravitationally bound sets of particles and cells \citep{Springel2001,Dolag2009}. We use the terms ``subhalo'' and ``galaxy'' interchangeably even though, in general, \subfind objects may contain no stars and-or gas whatsoever (that is, they may be composed entirely of dark matter). Typically, the most massive subhalo within a cluster is the ``main'' or ``primary subhalo,'' and is called the ``central galaxy'' or ``brightest cluster galaxy;'' all other subhalos may then be considered ``satellite galaxies (satellites)'' or ``cluster members,'' although at times we consider all other galaxies within a given aperture as satellites (see \S~\ref{sec:meth_sample} for more details). We follow the evolution of galaxies using \sublinkgal, which constructs merger trees for subhalos by searching for descendants with common stellar particles and star-forming gas cells \citep{Rodriguez-Gomez2015}. 

Throughout, by intracluster medium (ICM) gas, we mean all FoF gas within the cluster-centric aperture $[0.15,\,1.0]\rvir$, irrespective of temperature, excluding gas gravitationally bound to satellites, but including gas bound to the BCG. By construction, we also exclude all gas that belongs to other nearby halos. We check that our results remain qualitatively similar when using instead only gas that is gravitationally bound to the BCG. The adopted BCG-ICM boundary at $0.15\rvir$ ensures that we do not include extended cool interstellar medium (ISM) gas in our measure of the cool ICM. Further, at distances outside the ICM-IGM boundary $>\rvir$, the cool gas mass tends to decrease rapidly. This remains true when including all gas in the entire zoom simulation, that is, beyond the FoF membership. We expand on the spatial extent of the cool gas in \S~\ref{sec:results_spaceandtime}. 

We consider cool gas as all gas with temperatures $\leq 10^{4.5}$~K, which, by definition and construction in our galaxy formation model, includes all star-forming gas. We also use the terms ICM and BCG at all cosmic times, while we may refer to the cluster in its entirety as a ``cluster progenitor'' or ``protocluster'' at redshifts $z>0$.

\begin{figure*}
    \includegraphics[width=\textwidth]{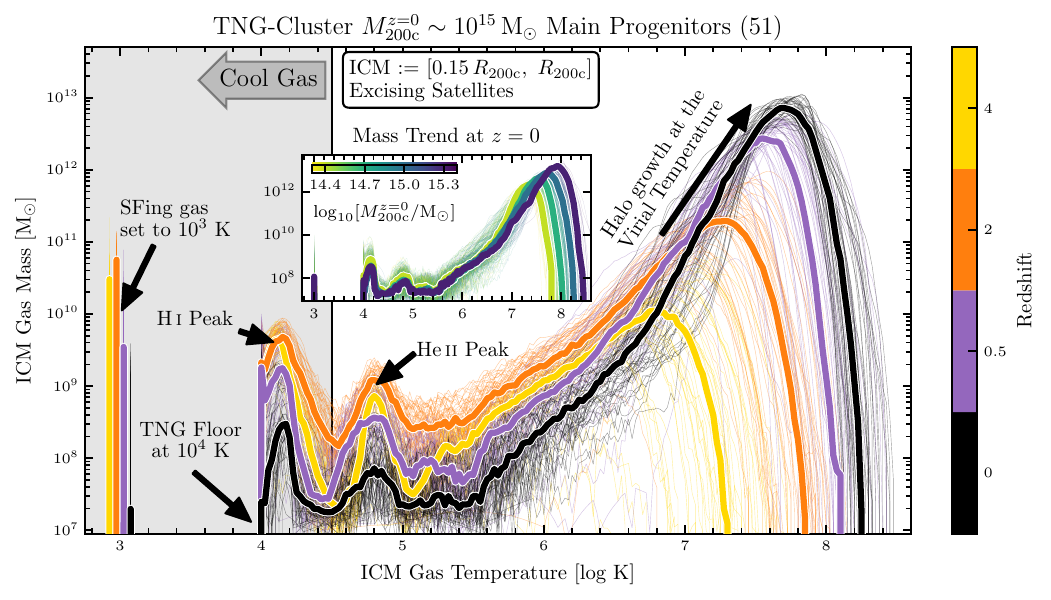}
    \includegraphics[width=\textwidth]{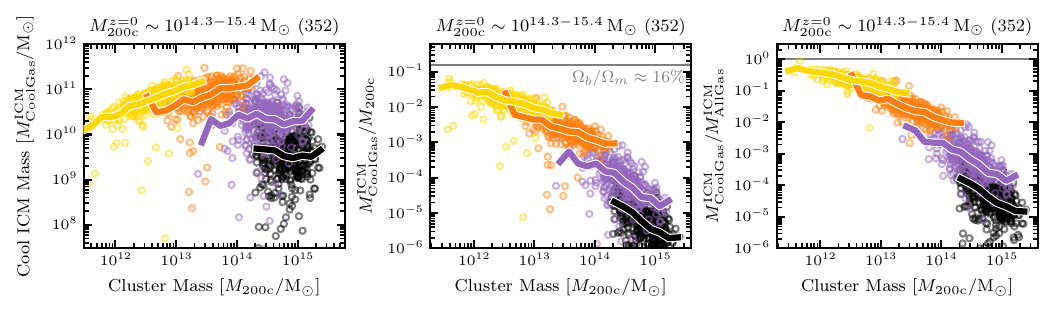}
    \caption{
    {\bf The evolution of the ICM temperatures, cool gas masses, cool ICM to total halo, and cool ICM to total ICM mass fractions in TNG-Cluster since $\mathbf{z=4}$.}
    The ICM is all FoF gas in the aperture $[0.15,\,1.0]\rvir$, excising satellites. Cool gas has temperatures $\leq 10^{4.5}$~K.
    \textit{Main Panel}: The distribution of ICM temperatures of each cluster of $z=0$ mass $\sim10^{15}\,\msun$ (51 clusters) as thin curves and the median of this sample as thick curves, where the color denotes the redshift. In the inset we show how the ICM temperature distribution varies with cluster mass today. We label prominent features in the ICM temperature distribution (see text for details; we offset the star-forming gas temperatures slightly for visibility).
    \textit{Bottom Panels}: The cool ICM mass (left), cool ICM to total cluster (center), and cool ICM to total ICM (right) mass fractions as functions of cluster mass and redshift (color) for all 352 clusters (and cluster-progenitors) in TNG-Cluster. We plot each cluster as circles and the median trend with mass as thick curves, colored by redshift.
    According to TNG-Cluster, the ICM of cluster progenitors were cooler, having more total cool gas, cooler average temperatures, and a larger ICM mass fraction in cool gas.
    }
    \label{fig:ICMGasEvolution}
\end{figure*}


\section{The cooler past of the ICM} \label{sec:results}

\subsection{The ICM was cooler in the past} \label{sec:results_coolpast}

In Fig.~\ref{fig:ICMGasEvolution} we show the evolution of the ICM temperatures: according to TNG-Cluster, the cluster progenitors had more cool gas, larger cool ICM to total halo mass fractions, and larger cool ICM to total ICM mass fractions than their descendants today. 

More specifically, in the main panel of Fig.~\ref{fig:ICMGasEvolution}, for each cluster of present-day mass $\sim10^{15}\, \msun$ (51 clusters), we plot the gas mass distribution of ICM temperatures as thin curves, colored by redshift. We include the cluster-wide medians at each redshift as thick curves. All FoF gas in the aperture $[0.15\rvir, \rvir]$, excluding satellites, constitutes the ICM (see \S\ref{sec:meth_sample}), and we define cool gas to be all gas with temperatures $\leq 10^{4.5}$~K (\S\ref{sec:meth_sample}), denoted by the shaded region.  

Today at $z=0$ for clusters of mass $\mvir\sim10^{15}\,\msun$, the majority of the ICM is hot at temperatures $\approx 10^{7.5-8}$~K ($kT \approx 3-9$~keV), approximately at the virial temperature and primarily heated via shocks when the gas was accreted \citep[][]{White1978}. The position of this peak at the virial temperature depends on the cluster mass $\tvir\propto \mvir^{2/3}$. As the halo mass grow, so do its gravitational potential well and the amount of gravitational potential energy that accreting gas converts into thermal energy. In the inset, we show how the $z=0$ ICM temperature distribution varies with cluster mass. The virial temperature and gas mass at this temperature increase with cluster mass, as expected.

At super-virial temperatures $\gtrsim 10^8$~K, the gas radiates efficiently via free-free bremsstrahlung emission. The gas mass at these temperatures decreases exponentially. The other features of the ICM temperature distribution depend largely on the cooling function, which is only indirectly related to halo mass. At sub-virial temperatures $\sim10^{5-7}$~K, in addition to bremsstrahlung radiation, the gas can cool via recombination and metal cooling, where the relative importance of the latter increases with increasing gas metallicity and decreasing halo mass. At cooler temperatures there are two peaks in the distribution at $\approx10^{4.2},\ 10^{4.8}$~K where cooling via bound-bound collisional excitation of atomic Hydrogen \ion{H}{i} and singly ionized Helium \ion{He}{ii} dominate the cooling function respectively. Then, due to the TNG cooling floor at temperature $10^4$~K, there is no ICM of lower temperatures, except for the star-forming gas, which we manually place at $10^{3}$~K (offset slightly for visibility; see \S~\ref{sec:meth_tng}, \ref{sec:disc_sf} for details). 

At higher redshifts, along the progenitors of the simulated clusters, the position of the hot, virial temperature peak moves to cooler temperatures, partly because the halo masses were smaller in the past. The relative amplitude of this peak, however, decreases due to both halo growth and an increased ability for cool gas to exist in the ICM. Despite the cluster progenitors having on average lower ICM metallicities (not shown), a larger fraction of their ICM is at sub-virial temperatures ($\lesssim 10^{7}$~K) in comparison to their descendants, implying that gas is cooling more efficiently. This increased cooling efficiency at higher redshift comes from the average increased density of Universe at earlier times, as the gas cooling time decreases with density -- we extensively expand upon this in \S~\ref{sec:why_cooling}. Additionally both the absolute and relative amplitudes of the peaks at \ion{H}{i} and \ion{He}{ii} increase with redshift. 

Is a dependence of cool ICM mass ($\mcgasicm$) with cluster mass $\mvir$ and redshift expected? In the bottom left panel of Fig.~\ref{fig:ICMGasEvolution}, we plot all 352 clusters (circles) and the median trend with mass (thick curves) at four example redshifts (color). For the $z=0$ clusters, there is little to no trend in $\mcgasicm$ with cluster mass, and  the clusters host on average $\sim 10^{9.5}\, \msun$ of cool ICM. At $z=0.5$, there is again an approximately flat trend with mass, but the normalization is now higher at $\sim10^{10}\, \msun$. At higher redshifts $\gtrsim2$, the cool ICM mass increases with halo mass. Between redshifts $\sim2-4$, the power law index (slope in the log-log plot) of $\mcgasicm$ as a function of $\mvir$ remains approximately constant, but the normalization still increases with redshift. That is, while a protocluster of mass $\mvir\sim10^{13}$ at $z\approx2$ has $\sim10^{10.5}\, \msun$ of cool gas in the ICM, at $z\approx4$ a similar mass cluster has $\sim10^{11}\, \msun$ of cool ICM. We speculate that at these early times while the BCGs are largely still star-forming, their gaseous atmospheres have not yet been significantly affected by kinetic mode feedback from the central SMBH, which could either heat up cool or prevent the cooling of hot ICM; see \S~\ref{sec:why_SMBH} for a discussion on this. 

The cool ICM to total halo fraction is quantified in Fig.~\ref{fig:ICMGasEvolution}, bottom center, where we include the global baryon fraction $\Omega_b / \Omega_m \approx 0.16$ for reference (gray line; constant with redshift). At all redshifts and masses considered, the cool ICM fraction decreases with cluster mass. At low redshift $\lesssim0.5$, the flat cool ICM mass trends with cluster mass (bottom left) translate to power law indices of the cool ICM fraction trends of $\approx -1$, where the normalization also decreases with redshift. That is, a cluster of mass $\sim10^{15}\, \msun$ at $z=0$ has a cool ICM to total halo fraction of $\sim10^{-5.5}$, while at $z=0.5$ a similar mass cluster has a fraction of $\sim 10^{-5}$. At higher redshifts $z\gtrsim2$, slopes of the cool ICM fraction trends are still negative but flatter than those at lower redshifts, a reflection of the positive cool ICM mass trend at these redshifts (bottom left). The slopes are similar at redshifts $\sim2-4$, but the normalization increases with redshift. A protocluster of mass $\sim10^{13}\, \msun$ at redshift $\sim2$ has a cool ICM to total halo fraction of  $\sim 10^{-2.5}$, while a similar mass cluster at $z=4$ has a fraction of $\sim 10^{-2}$. The lowest mass protoclusters considered here -- $\mvir\sim10^{11.5}\, \msun$ at redshift $\sim4$ -- have cool ICM to total halo fractions $\sim10^{-1}$, meaning that $\approx60$~per~cent of the baryons within the cluster are in the cool ICM.

Lastly we examine the fraction of ICM that is cool (Fig.~\ref{fig:ICMGasEvolution}, bottom right), where we mark where the fraction is one, where the entire ICM is cool (gray line). At all redshifts, the cool to total ICM mass fraction decreases with halo mass with similar power law indices, but the normalizations increase with redshift. A cluster of mass $\sim10^{15}\, \msun$ today has only $10^{-5}$ of its total ICM in the cool phase, while a similar mass cluster at redshift $\sim0.5$ has $\sim10^{-4}$ of its ICM in the cool phase. The ICM in the lowest mass protoclusters considered here at $\sim10^{11.5}\, \msun$ at redshift $\sim4$ is $\approx50$~per~cent cool. 

To summarize, according to TNG-Cluster, at a fixed cluster mass, clusters had more cool gas in their ICM at earlier times. While smaller clusters have cooler virial temperatures and a higher fraction of ICM in the cool phase, the conditions at higher redshifts were more conducive for either cooling hot ICM and-or maintaining cool ICM.

\subsection{The cool cluster gas across space and time} \label{sec:results_spaceandtime}

\begin{figure}
    \includegraphics[width=\columnwidth]{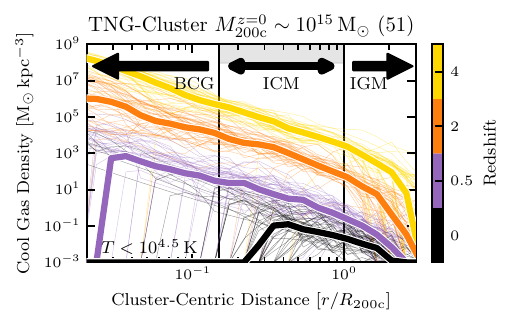}
    \caption{
    \textbf{The evolution of the cool gas radial profiles in TNG-Cluster since $\mathbf{z=4}$.}
    We plot the cool gas density radial profiles of the 51 clusters with present-day mass $\sim10^{15}\, \msun$ (thin curves) and the medians of this sample (thick curves), where the color denotes redshift. We normalize the radial profiles by the virial radius $\rvir$ at that redshift. At all redshifts and cluster-centric distances considered, the cool gas density increases with redshift. Outside of the adopted ICM-IGM boundary at $\rvir$, the cool gas densities begin to drop exponentially.
    }
    \label{fig:coolgasprof_evolution}
\end{figure}

\begin{figure*}
    \includegraphics[width=\textwidth]{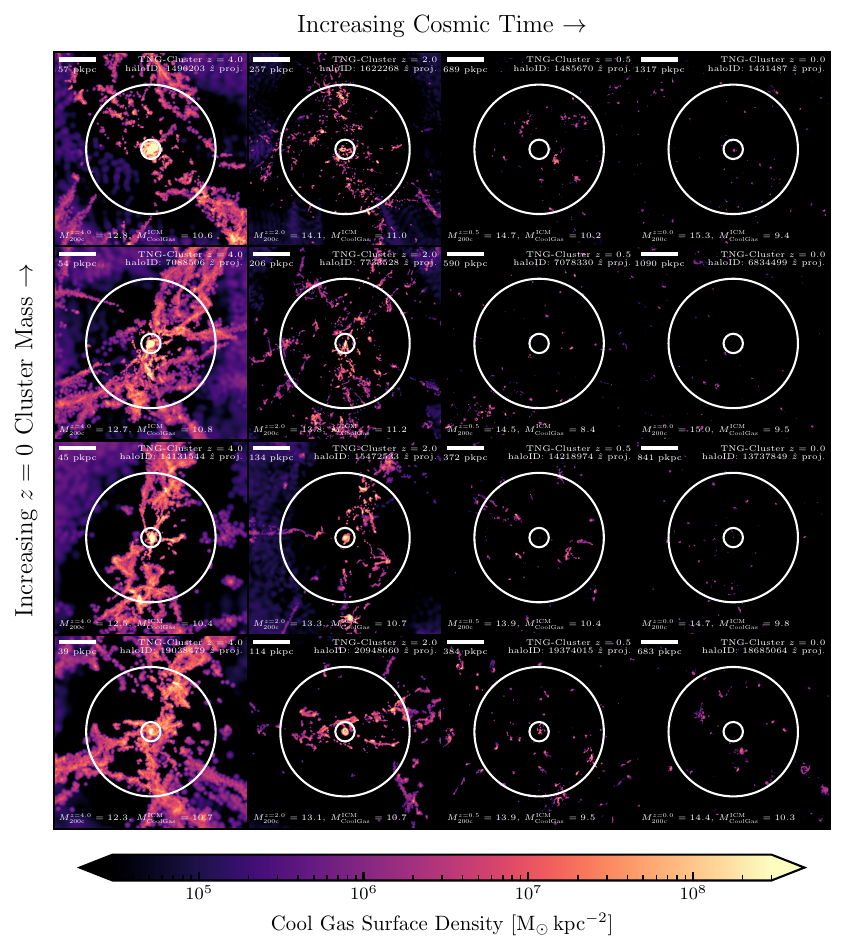}
    \caption{
    {\bf Examples of the cool gas surface density as functions of cosmic time and cluster mass in TNG-Cluster systems.}
    Each panel shows the total cool gas (temperature $\leq 10^{4.5}$~K) surface density, including the BCG and potential satellites, within a cube of $3\rvir$ centered on the BCG, where we project the gas cells using a cubic spline of variable kernel size according to the gas cell size. The white circles mark the adopted BCG-ICM and ICM-IGM boundaries at $0.15\rvir, \rvir$, and the scale in the upper-left is $0.5\rvir$ in size. We include the halo ID at the given redshift in the top right, and the halo $\mvir$ and cool ICM mass $\mcgasicm$ at each redshift in the lower-left in units of $[\log_{10}\, \msun]$. Each row shows the evolution of an individual cluster, and each column shows the mass trend at a fixed redshift. According to TNG-Cluster, at all halo masses, clusters tends to have less cool gas today than at earlier times.
    }
    \label{fig:CoolGasSurfaceDensity_mosaic}
\end{figure*}

\begin{figure}
    \includegraphics[width=\columnwidth]{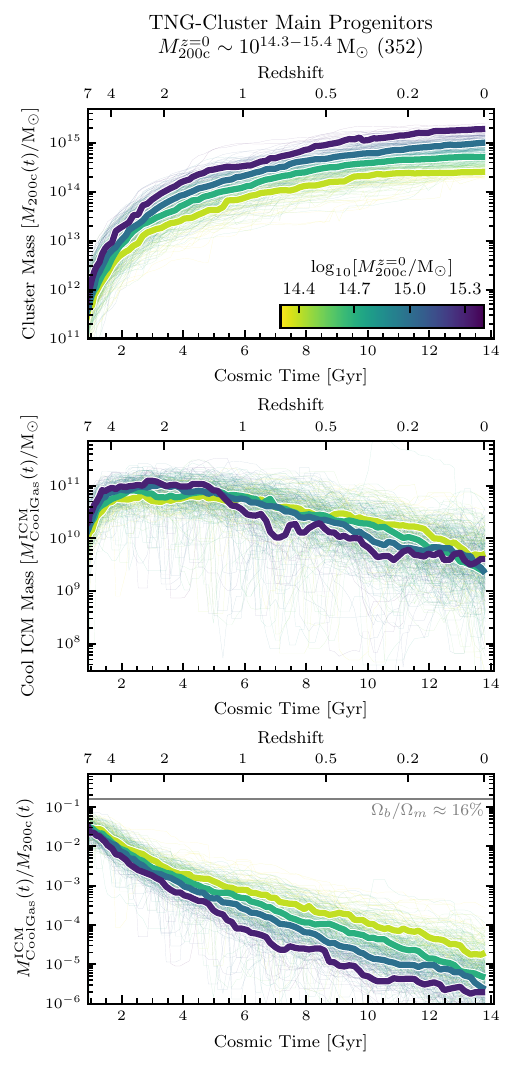}
    \caption{
    \textbf{The evolution of individual clusters and their cool ICM since $\mathbf{z=7}$.}
    For all 352 clusters in TNG-Cluster we plot the evolution of the cluster mass $\mvir$ (top panel), cool ICM mass $\mcgasicm$ (center panel), and cool ICM to total halo mass fraction (bottom panel) as thin curves colored by their $z=0$ cluster mass, and we include the median evolutionary trends at a fixed $z=0$ cluster mass (thick curves). The cool ICM mass and cool ICM to total halo fraction curves are averaged over five snapshots, corresponding to $\approx 750$~Myr. While the cool ICM mass and cool ICM to total halo fraction for individual clusters (center and bottom panels) evolve in a complicated manner, increasing or decreasing by more than an order of magnitude at times, the median trends show an average decrease in cool ICM mass and cool ICM to total halo fraction since $z\lesssim4$, according to TNG-Cluster.
    }
    \label{fig:ICMCGM_MPBs}
\end{figure}

\begin{figure}
    \includegraphics[width=\columnwidth]{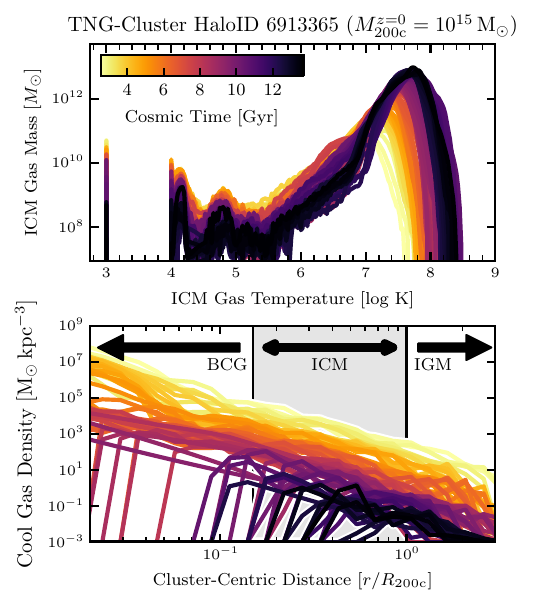}
    \caption{
    \textbf{Evolution of the ICM temperatures and external cool gas profile for an example halo from TNG-Cluster.}
    We plot the evolution of the ICM temperature distribution (top panel) and external cool gas radial profile (bottom panel) for an example cluster of $z=0$ mass $\sim10^{15}\, \msun$. Over the past $\approx10$~billion years, the hot ICM mass near the virial temperature increases while the cool gas density decreases at all radii, but with more evident suppressions proceeding inside-out.
    }
    \label{fig:example_evolution}
\end{figure}

We show in Fig.~\ref{fig:coolgasprof_evolution} the cool gas density radial profiles, normalized by the virial radius $\rvir$ at the given redshift (see Appendix~\ref{app:normalizations} for different normalizations). Here we show clusters of $z=0$ mass $\sim10^{15}\, \msun$ (51 clusters, thin curves; see Appendix~\ref{app:halo_mass} for the trend with halo mass today) and the median of this sample at each redshift (color). We label the regions of interest and their definitions for this work: BCG at distances $<0.15\rvir$; ICM in the aperture $[0.15\rvir,\, \rvir]$; and IGM at distances $>\rvir$. These present-day clusters have on average little cool gas in their BCGs, reflecting the lack of cool gas available to, for example, form stars: a majority of the BCGs in TNG-Cluster are quenched at $z=0$ \citep[][see fig.~11 and discussion therein]{Nelson2024}. That is, only $\approx10$~per~cent of BCGs are star-forming today, according to the standard criterion of specific star formation rate (sSFR) $>10^{-11}$~yr, which is broadly consistent with observations of South-Pole-Telescope-selected clusters \citep{McDonald2016} and would be expected since a majority of BCGs in TNG-Cluster contain no cool gas today. Within the TNG model, kinetic feedback from the SMBH is understood to remove the cool, star-forming ISM, to ultimately quench massive central galaxies, and to keep them quenched \citep[e.g.,][]{Pillepich2018,Nelson2018,Weinberger2018,Terrazas2020,Zinger2020}, and this is the case for TNG-Cluster as well.
In the cluster outskirts at $r\gtrsim\rvir$, the cool gas density begins to drop exponentially. We note that cool gas profile still drops exponentially when including all gas in the simulation. 

At higher redshifts, the cool gas density at a fixed cluster-centric distance increases at all distances. Especially at redshifts $\gtrsim2$, there is a significant amount of cool gas within the BCG at distances $<0.15\rvir$, agreeing with expectations that the cluster cores tend to be more cool-cored with increasing redshift \citep{Lehle2024}. Additionally, the cool gas densities are higher than what would be expected from the cosmic evolution, which scales as $\propto (1+z)^3$, affirming that the conditions for cool gas are more conducive at higher redshifts. These results motivate our fiducial definition of the ICM: all FoF gas in the aperture $[0.15\rvir,\, 1.0\rvir]$ that is not bound to satellites. We note that in general, the ICM extends beyond the virial radius. Especially at later times $z\lesssim1$, the shock radius is likely located beyond the virial radius \citep[e.g.,][]{Birnboim2003,Voit2003,Zinger2018a}, and satellite stripping likely begins outside of the virial radius \citep[e.g.,][]{Bahe2013,Ayromlou2021b,Rohr2023,Zinger2024}.

Studying the evolution of individual objects allows us to better understand the population trends across redshift. In Fig.~\ref{fig:CoolGasSurfaceDensity_mosaic} we show the cool gas surface density for four example individual clusters (rows) at four example redshifts (columns). Each image includes all cool gas, including the BCG and satellites, within a cube of $3\rvir$ centered on the BCG, where we project the gas surface density using a cubic spline of variable kernel size according to the gas cell size. The white circles mark the BCG-ICM and ICM-IGM boundaries at $0.15\rvir,\, \rvir$ respectively, and the scale in the upper-left is $0.5\rvir$ in size. For each of the example clusters, their high redshift $z\gtrsim 2$ progenitors had more cool gas than their descendants today. Especially at $z=4$, large scale cool filaments are present and directly feeding cool gas into the ICM, where no such cool filaments are present at later times $z\lesssim 0.5$, as expected \citep[e.g.,][]{Zinger2016,Birnboim2016}. Many gaseous satellite galaxies are visible at all redshifts, but at $z\sim4$ many satellites are co-spatial with the cool filaments. The BCGs at these high redshifts are largely still star-forming, and there are morphological signs of cool gas disks in these protoclusters. At a fixed time, the cool gas maps have qualitatively similar morphologies across cluster masses. At $z\gtrsim2$, the total cool ICM, excluding satellites, increases with cluster mass, while at later times there is no trend with cluster mass (see also \S~\ref{sec:results_coolpast}). 

From individual systems to the whole population, Fig.~\ref{fig:ICMCGM_MPBs} shows the evolution of the cluster mass $\mvir$ (top panel), cool ICM mass $\mcgasicm$ (center panel), and cool ICM to total halo mass fraction (bottom panel) since $z=7$ of all 352 clusters (thin curves) colored by their $z=0$ mass. The median trends at a fixed mass are given with thick curves. The cool ICM mass and cool ICM to total halo fraction curves are averaged over five snapshots $\approx 750$~Myr. While the evolution of individual cluster masses (top panel) may be complex, involving discrete jumps in mass likely due to merger events, the median trends in TNG-Cluster smoothly and monotonically increase with time. Generally more massive clusters today tend to be more massive at all times. Therein, more massive $z=0$ clusters tend to reach a given characteristic halo mass at earlier times; for an example characteristic halo mass at $10^{13}\, \msun$, a massive cluster of mass $\sim10^{15.3}\, \msun$ today was already at this mass by redshift $\approx4$, while a less massive $z=0$ cluster of mass $\sim10^{14.3}\,\msun$ reached this mass at $z\approx3$, $\sim1$~Gyr later. 

The evolution of the cool ICM mass (center panel) and cool ICM to total halo mass fraction (bottom panel) are more complex. At early times $z\gtrsim4$, clusters were gaining more cool gas with time. At $z\sim2-4$ the average cool ICM mass is maximum for all clusters, interestingly corresponding with the peak of cosmic star formation. At a fixed time here ($z\sim2-4$), more massive protoclusters have more cool gas in their ICM than lower mass ones (see also Fig.~\ref{fig:ICMGasEvolution}, bottom left panel). At $z\lesssim2$, the cool ICM masses tend to decrease with time. We note that the median curve of the most massive clusters today ($\mvir\sim10^{15.3}\, \msun$) is lower than, but within the scatter of, the trends for the other mass bins, and this population has the lowest number statistics. In individual systems, the evolution of the cool ICM may be much more complicated. Many clusters show large jumps the cool ICM mass and thereby the cool ICM to total halo mass fraction (bottom panel) by one to two orders of magnitude, before potentially returning to the average value. For the cool ICM to total halo mass fraction (bottom panel), the median fractions decrease with time since $z\sim4$. At all times since $z\sim4$, but especially so at later times $z\lesssim2$, more massive (proto) clusters have lower cool gas to total mass fractions (see also Fig.~\ref{fig:ICMGasEvolution}, bottom center). 

Lastly in Fig.~\ref{fig:example_evolution}, we demonstrate how the distribution of ICM temperatures (top panel) and the cool gas density radial profile (bottom panel) evolve for an individual system. With cosmic time, both the amplitude and the position of the hot gas peak near the virial temperature increase. Simultaneously the ICM mass decreases at all cool temperatures $\lesssim10^{4.5}$~K. At all radii considered, the cluster exhibits less cool gas with cosmic time, and more prominently so in the inner regions of the ICM or BCG. The cool gas density profiles approximate power laws in the BCG and ICM at earlier times, before declining exponentially in the IGM. With increasing cosmic time, the cool gas density decreases to below the TNG-Cluster resolution limit and it does so in an inside-out fashion: this suggests that processes in the BCG may be crucial for determining the cool ICM content.

\section{Why was there more cool gas in the past?} \label{sec:why}

\begin{figure*}
    \centering
    \includegraphics[width=\textwidth]{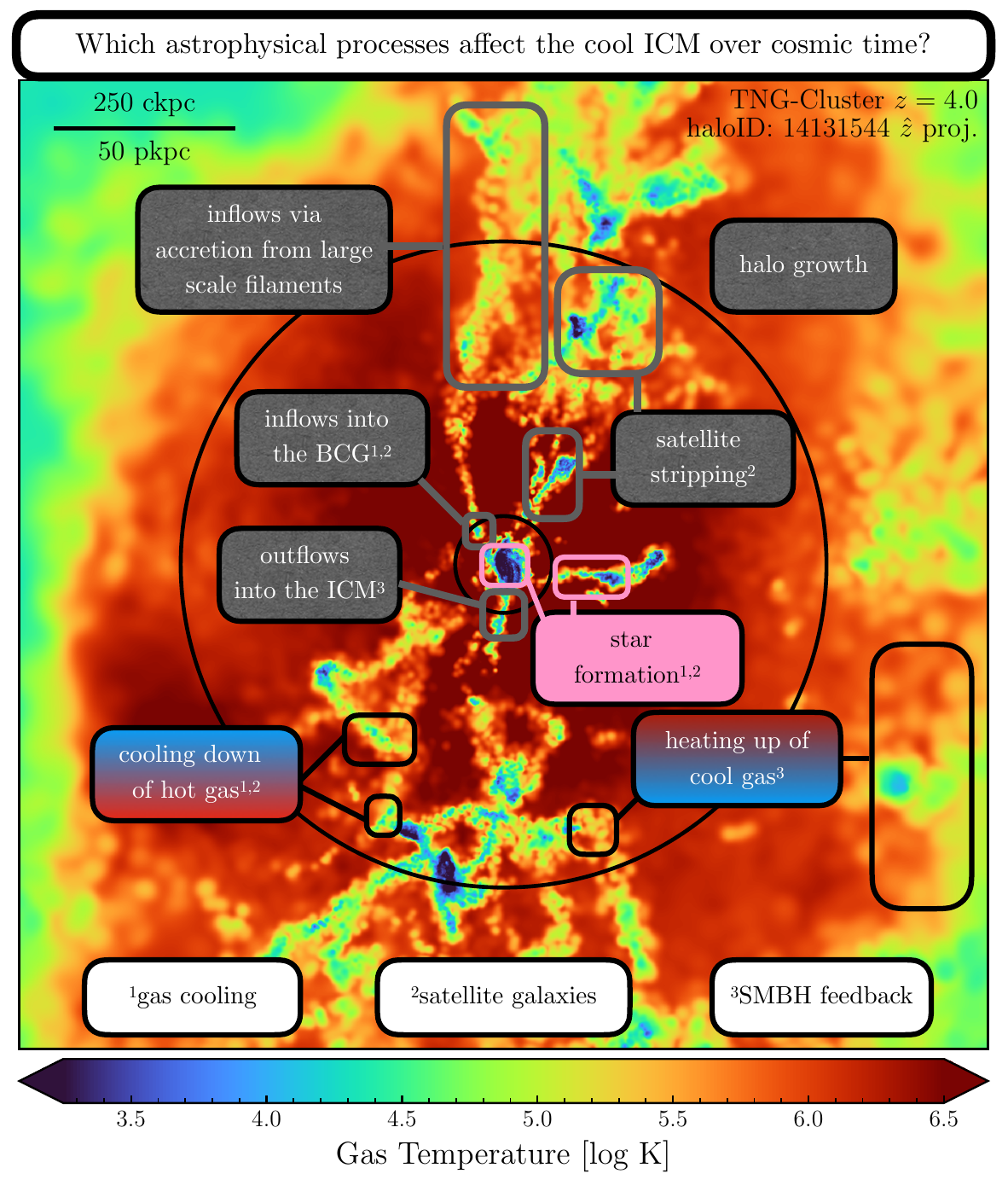}
    \caption{
    \textbf{Schematic detailing the physical mechanisms responsible for the amount of cool ICM.}
    We plot the mass-weighted temperature map of an example protocluster at $z=4$, including all gas within a cube of $3\rvir$ using a cubic spline kernel. The annotations are as in Fig.~\ref{fig:CoolGasSurfaceDensity_mosaic}.
    We label the various ways -- namely fluxes at a fixed gas temperature (gray) and changes of gas phase at a fixed spatial location (blue-red and pink) -- in which the cool ICM mass can change and ascribe the dominant astrophysical phenomena (bottom, numbered) responsible for either the production, destruction, or survival of cool gas on non-cosmological time scales. Many of these processes are interconnected and their relevant importance change with redshift; see text for details.
    }
    \label{fig:schematic}
\end{figure*}

The growth and evolution of galaxy clusters is the result of hierarchical structure formation convolved with baryonic physics and galactic feedback. In Fig.~\ref{fig:schematic} we visualize the complexity and inter-connectedness of these various processes with the goal of understanding not only how the cool ICM mass changes with time, on non-cosmological time scales, but also why it ubiquitously decreases over the past $\sim10$~billion years (see also \citealt{Peroux2020a,Donahue2022} for recent reviews of the cosmic baryon cycle). 

In particular, in Fig.~\ref{fig:schematic}, we plot the mass-weighted temperature map of an example protocluster at $z=4$ from TNG-Cluster, including all gas within a cube of $3\rvir$ centered on the BCG and projected using a cubic spline kernel, where the annotations are as in Fig.~\ref{fig:CoolGasSurfaceDensity_mosaic}. Throughout the cluster and its environment there are complex, multiphase structures. In the following, we list and describe sources and sinks of cool gas in the ICM, noticing that its amount can change due to both mass fluxes across the ICM ``boundaries'' (at fixed cool gas temperature) as well as changes in temperature (at a fixed spatial location). Once we identify the physical processes that may affect the production, destruction, or survival of cool gas in the otherwise hot ICM, we assess whereas, and how, the prevalence of such individual processes may change across cosmic epochs as clusters assemble and evolve.\\

\textit{Halo growth and cool ICM.} 
First, halo growth is disfavorable for the cool ICM. As shown in Figs.~\ref{fig:ICMGasEvolution},~\ref{fig:ICMCGM_MPBs},~and~\ref{fig:example_evolution}, the amount of cool ICM decreases as clusters grow, because of their growth in mass. The mode of the temperature distributions at the virial temperature increases with halo mass and time, which then increases the temperature contrast between the volume filling hot gas and the cooler gas clouds: this decreases the chances of cool gas survivability \citep[e.g.,][]{Sparre2020,Fielding2022}.\\

\textit{Mass fluxes of cool gas in the outer regions.}
Clusters do not grow in isolation nor in the absence of galactic feedback. Therefore, both the cosmological growth of structure and galaxy evolution processes can potentially affect the ways in which the cool ICM mass change via cool gas fluxes at the ICM spatial boundaries. Here we firstly consider the ICM-IGM boundary. Large scale filaments, which are visible in the figure as elongated structures of cool gas (see also Fig.~\ref{fig:CoolGasSurfaceDensity_mosaic}, especially the left-most columns), can channel cool gas from the IGM into the ICM, at times even directly into the BCG. As described in the introduction, the clusters likely grew by accreting cool gas at high redshifts $\gtrsim2$ and hot gas at lower redshift $\lesssim2$.
That is, this source of cool gas at high redshifts is much less prominent at lower redshifts, where clusters progenitors at $z=2$ had $\approx100\times$ higher cool gas accretion rates than their descendants today, according to TNG-Cluster (see also Fig.~\ref{fig:CoolGasSurfaceDensity_mosaic}, right-most columns). According to our analysis of the simulated systems, many satellite galaxies are also located co-spatially to these filaments, and in general, it is expected that satellites tend to accrete along filaments rather than spherically \citep[e.g.,][]{Fielding2020b,Kuchner2022}. We return to the role of satellite galaxies below in \S~\ref{sec:why_satellites}. In the other direction, it is possible that cool gas can leave the ICM due to, for example, outflows driven by the SMBH or stellar feedback originating within the BCGs. At $z=0$, the clusters have little outflowing gas into the ICM, and the outflowing gas that does exist tends to be hot \citep{Ayromlou2023,Ayromlou2024}. At lower host masses both today and in the past for the cluster progenitors, the stellar- and SMBH-driven outflows in the TNG model can extend beyond $\rvir$, but the outflows tend to be shock heated \citep{Weinberger2017,Nelson2019b,Pillepich2021}. We check that the cool gas outflow rates at $\rvir$ are on average $\lesssim 10^{-1}\, \msun\, \mathrm{yr^{-1}}$ at all studied cluster masses and redshifts in TNG-Cluster. \\

\textit{Mass fluxes of cool gas in the inner regions.}
According to TNG-Cluster and the TNG model in general, there can be a cool gas flux also at the BCG-ICM boundary at $0.15\rvir$ \citep[e.g.,][]{Nelson2019b}. Feedback from the SMBH in kinetic mode and from supernovae in the form of galactic winds can drive cool gas outflows out of the innermost galaxy regions. However, within the TNG model, as the gas is pushed outwards, it also heats up via shocks before reaching the ICM \citep{Weinberger2017, Pillepich2021}. Conversely, cool gas may return to the BCG via galactic fountains or precipitation onto the BCG, acting as fuel for star formation \citep{Fraternali2008,Voit2015}. A majority of the BCGs of TNG-Cluster are still star-forming at high redshift $z\gtrsim3$, while the opposite is true today \citep{Nelson2024}. Thereby, while some cool gas falls onto the BCG at early times, this becomes negligible at later times, at least partially contributing to the decrease in the cool ICM mass since $z\lesssim 4$.\\

\textit{In-situ changes of gas temperature.}
Within the ICM itself -- assuming no gas fluxes -- gas can change phase in three distinct ways: (i) hot gas can cool down; (ii) cool gas can heat up; or (iii) cool gas can form stars. We return to the ICM in-situ star formation in \S~\ref{sec:disc_sf}. Gas cooling and heating are affected by multiple factors: as halos and their virial temperatures grow with time, it becomes both more difficult for cool gas to survive in the hotter environments and more difficult for hot gas to cool down. In fact, hot gas may cool into cold clouds \citep[e.g.,][]{sharma2012a,Sharma2012b,Voit2017}. That is, a cool phase may condensate out of a hot medium via thermal instabilities, facilitated by density perturbations, which in turn may eventually rain down onto the BCG \citep[][]{Voit2015,Voit2021}. In a cosmological context, density perturbations could be caused by the passage of satellites \citep[][]{Nelson2020, Fielding2020b,Ramesh2023b}, and satellites can directly deposit their cool ISM into the ICM via ram pressure stripping \citep[e.g.,][]{Nelson2020,Rodriguez2022,Rohr2023,Saeedzadeh2023}. At the same time, in a galaxy evolution context, feedback from the central SMBH of the BCG (or other massive cluster galaxies) can both directly heat up cool gas (see above) or increase the cooling time of the hot ICM \citep{Truong2020,Zinger2020}. If the SMBH feedback causes local overdensities, then it may trigger localized cooling, despite the overall effect leading to outflows and heating \citep{Zinger2020}. \\

In the remainder of this analysis, we use TNG-Cluster to further and explicitly quantify the thermal instability cooling framework in the ICM in a full cosmological context (\S~\ref{sec:why_cooling}) and to evaluate and highlight the role of satellites and the effects of SMBH on the cool ICM (\S~\ref{sec:why_satellites} and \S~\ref{sec:why_SMBH}, respectively).


\subsection{The decreasing importance of ICM cooling towards \texorpdfstring{$z=0$}{z=0}} \label{sec:why_cooling}

\begin{figure*}
    \includegraphics[]{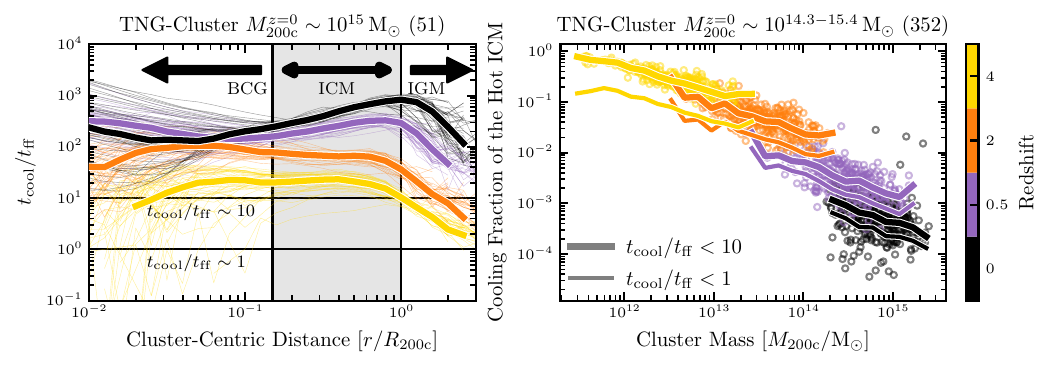}
    \caption{
    \textbf{The ICM in TNG-Cluster progenitors was more prone to cooling in the past.}
    We examine the ability of the hot ICM (in the aperture $[0.15\rvir, \rvir]$ with temperatures $>10^{4.5}$~K, excising satellites) to cool, using the canonical criterion of cooling to free-fall ratio $\tcool / \tff < 10$ as a proxy for cooling gas. 
    \textit{Left panel}: We plot the radial azimuthally averaged profile of the cooling to free-fall time for the 51 systems from TNG-Cluster with $z=0$ mass $\sim10^{15}\, \msun$ (thin curves) and the medians of this sample (thick curves), where the color denotes redshift. We normalize the radial profiles by the virial radius $\rvir$ at that redshift. We include horizontal lines at the classical (catastrophic) cooling to free-fall ratios $\tcool / \tff \sim (1)\ 10$. The average cooling to free-fall ratio in the ICM increases with cosmic time, meaning that the ICM is less susceptible to cooling today than it was in the past.
    \textit{Right panel}: The mass fraction of the hot ICM that is susceptible to cooling, with cooling to free-fall time $<10$, for all clusters in TNG-Cluster (circles) as a function of cluster mass and redshift, where we include the median trend with mass as thick curves. We also mark the median trend of the fraction of ICM with cooling to free-fall time $<1$ as thin curves. At a fixed redshift, the cooling fraction of the hot ICM decreases with cluster mass; at a fixed cluster mass, the cooling fraction increases with redshift. 
    \label{fig:coolingtime}
    }
\end{figure*}

An important source of cool halo gas is the cooling of the ambient, volume-filling hot gas. When the cooling time $\tcool$ compared to the dynamical free-fall time $\tff$ drops below some threshold, the hot gas will tend to cool faster than it can be re-heated by the surrounding medium. Expressions for the cooling and free-fall times are
\begin{align}
    \tcool &= \dfrac{3}{2}\dfrac{(n_e + n_i)k_b T}{n_e n_i \Lambda_{\rm cool}} \\
    \tff &= \left(\dfrac{3\pi}{32G\bar{\rho}_{\rm tot}(<r)}\right)^{1/2} \approx \sqrt{2}\left(\dfrac{2r^3}{GM_{\rm tot}(<r)}\right)^{1/2}
\end{align}
where $n_e$, $n_i$ are the electron and ion abundances, $k_bT$ is the thermal energy, $\Lambda_{\rm cool}$ is the instantaneous net cooling rate, $G$ is the gravitational constant, $\bar{\rho}$ is the average total internal density, and $M_{\rm tot}(<r)$ is the total internal mass. Thereby, the cooling time is a local measurement that depends on the properties of the gas, whereas the free-fall time is spherically averaged and depends only on cluster-centric distance. A canonical criterion for gas cooling is $\tcool / \tff < 10$ \citep[e.g.,][]{sharma2012a,Voit2017}, whereas catastrophic cooling or cooling in the presence of gravity can occur when $\tcool / \tff < 1$ or $< 10$, respectively \citep{McCourt2012}. In fact, such a threshold can be higher in the presence of large local density perturbations \citep{Choudhury2019}.  In the following we consider gas fulfilling the  $\tcool / \tff < 10$ criterion to be susceptible to cooling, although in general, not all hot gas with $\tcool / \tff < 10$ may cool and not all hot gas with $\tcool / \tff > 10$ is thermally stable \citep[][]{Choudhury2019,Saeedzadeh2023}. 

Because the cooling time is proportional to the inverse of the gas density and the free-fall time to the inverse square root of the total density, gas becomes more susceptible to cooling if the local and-or the global density increases. The global density increases with redshift as $\rho \propto (1+z)^3$, meaning that in general, $\tcool / \tff \propto (1+z)^{-3/2}$ and global cooling is more efficient at higher redshifts. Additionally, the local density can be perturbed, for example, by the passage of satellites or by the presence of already existing over-dense, cool gas clouds \citep[e.g.,][]{Nelson2020,Fielding2020b,Ramesh2023b}. Perhaps these perturbations occurred more frequently at higher redshift (see \S~\ref{sec:why_satellites}). 

In Fig.~\ref{fig:coolingtime} we examine the ability for the ICM to cool across cosmic epochs. We plot the mass-weighted cooling to free-fall time radial profile for all TNG-Cluster systems with $z=0$ mass $\sim10^{15}\, \msun$ (51 clusters; left panel) as individual curves colored by redshift. We include the median curve at each redshift as thick curves and mark both the adopted ICM region and the canonical cool criteria. Here, we only consider hot gas with temperatures $>10^{4.5}$~K that is cooling $\Lambda_{\rm cool} < 0$. In the ICM, the average cooling to free-fall time decreases with redshift, showing that the higher-redshift ICM is in fact more susceptible to cooling, as expected. That is, at a cluster centric radius $\sim 0.5\rvir$, the average cluster today has cooling to free-fall time ratio of $\sim 10^{2.5}$ while a proto-cluster may have $\sim 20$, more than order of magnitude lower. However, the average ratios (averaged both across clusters and spherically within each cluster) are always (or typically) at $\tcool / \tff > 10$, suggesting that on average the ICM is in fact not cooling significantly. At lower host masses $\sim10^{12.5-14}\, \msun$ the cooling to free-fall time ratio in the CGM tends to still be $>10$, although there is gas at all radii fulfilling $\tcool / \tff < 10$, which tends to be infalling, \citep{Nelson2020}. At higher-redshifts $z\gtrsim 2$, the radial profiles are approximately flat or even decrease with cluster-centric distance, while the profiles at lower redshfits $z\lesssim 0.5$ increase with cluster-centric distance. In the ICM for present-day clusters, the cooling to free-fall time ratio tends to increase with cluster mass, while the ratio in the inner regions depends on whether the object is a cool core cluster \citep{Lehle2024}. 

The radial profiles of Fig.~\ref{fig:coolingtime} (left panel) are spherically averaged and represent the mass-weighted average of a possibly wide distribution of cooling to free-fall times. To probe the ICM that is susceptible to cooling, we compute the mass fraction of the ICM fulfilling the (catastrophic) cooling criterion $\tff / \tcool < (1)\ 10$ compared to the total hot ICM mass (Fig.~\ref{fig:coolingtime} right panel). We then plot this cooling fraction of the hot ICM as a function of cluster mass and redshift for all 352 clusters (circles) and include the median mass trend at a fixed redshift as thick curves (thin curves show the median trend of the fraction of the hot ICM fulfilling the catastrophic cooling criterion $\tcool / \tff < 1$). According to TNG-Cluster, at a fixed redshift, the cooling fraction of the hot ICM decreases with cluster mass, such that lower mass clusters likely have more cooling proportional to their total ICM. At a fixed cluster mass, the cooling fraction increases with redshift, agreeing with expectations. In fact, $\approx 100$~per~cent of the ICM in protoclusters of mass $\sim 10^{12}\, \msun$ at $z\sim 4$ are able to cool, and $\approx10$~per~cent may cool catastrophically. 

In summary, the ICM gas cooling was more significant in the past than it is today, at least partially explaining why the cool ICM mass decreases with time. What perturbed it to cool and what happens to it after cooling remains unanswered. We consider below one of the causes of these perturbations, satellite galaxies, and discuss this gas cooling to the point of forming stars in-situ in the ICM in \S~\ref{sec:disc_sf}.

\subsection{The decreasing importance with cosmic time of satellites as sources for cool ICM} \label{sec:why_satellites}

\begin{figure*}
    \includegraphics[width=\textwidth]{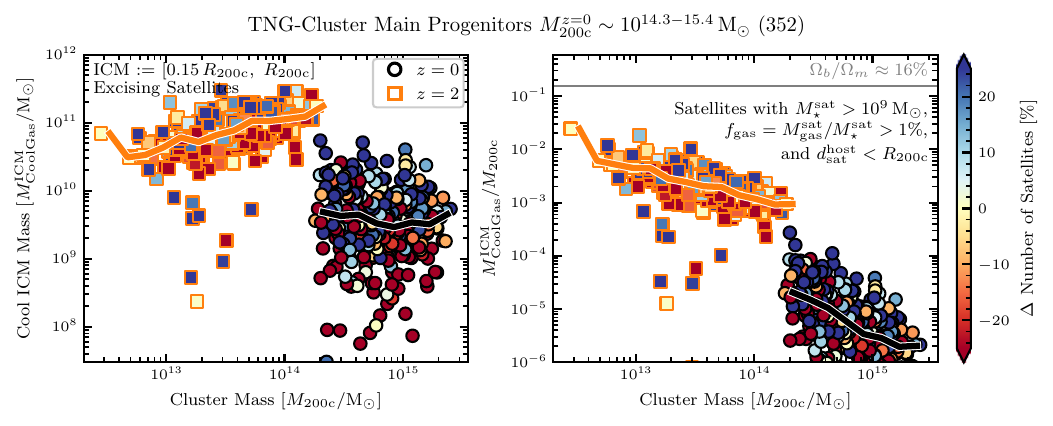}
    \includegraphics[]{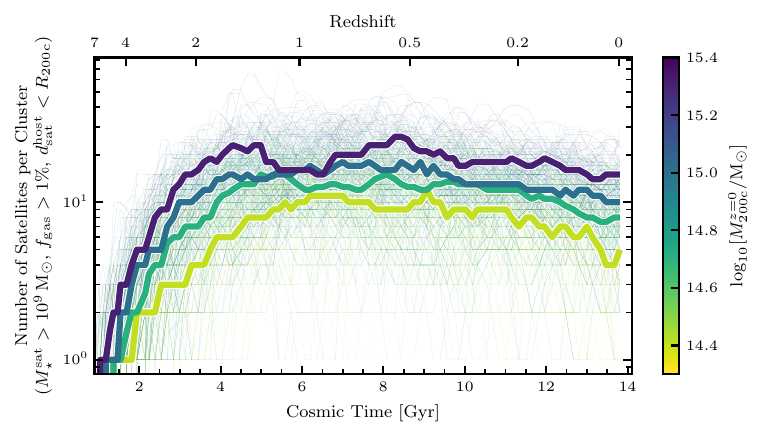}
    \caption{
    {\bf According to TNG-Cluster, at a fixed cluster mass and redshift, the cool ICM mass correlates with the number of gaseous satellites, and clusters tend to have fewer gaseous satellites today than in the past.}
    \textit{Top Panels}: At $z=0$ and 2 (black-outlined circles, orange-outlined squares), we demonstrate that at a fixed redshift the cool ICM mass increases with the relative number of satellites, that is, with the percentage difference between the number of satellites in a given cluster and the average number in a corresponding narrow bin of halo mass. Here, we only consider satellites with a stellar mass $>10^9\, \msun$, a gas to stellar mass fraction $>1$~per~cent, and within a cluster-centric distance $<\rvir$.
    \textit{Bottom Panel}: We show the evolution of the number of gaseous satellites per cluster over the past $\approx13$~billion years for all 352 clusters (thin curves), colored by their $z=0$ cluster mass (medians within a $z=0$ cluster mass bin as thick curves). Namely for all considered masses today, clusters have fewer gaseous satellites today than in the past at $z\approx1-2$, at least partially explaining why the cool ICM mass decreases with time.
    }
    \label{fig:ICMCGM_Nsatellites}
\end{figure*}

It is plausible that satellites can increase the cool ICM mass both via direct deposition of their cool ISM gas into the ICM via ram pressure stripping and by triggering hot gas to cool via density perturbations. The stripped gas is, in many cases, more metal rich than the ICM. This seeding of metals can also contribute to the enhanced cooling. Satellites are in principle also able to accrete cool gas from the surrounding medium, but in general, it is expected that the satellite accretion is negligible within the cluster environment \citep[][]{Larson1980,Balogh2000,vandeVoort2017,Rohr2023}. In the following, we quantitatively demonstrate that, in general, the cool ICM mass at a fixed cluster mass and redshift increases with the number of gaseous satellites.

In Fig.~\ref{fig:ICMCGM_Nsatellites}, we show that, at a fixed cosmic time, the number of gaseous satellites -- all galaxies with stellar mass $>10^9\, \msun$, gas to stellar mass fraction $>1$~per~cent, and within a cluster-centric distance $<\rvir$ -- increases with cluster mass (bottom panel). Therefore, to see trends of cool ICM mass with both the number of gaseous satellites and cluster mass (top panels), we compute the relative number of satellites per halo: within a narrow bin of halo mass at a fixed redshift, we compute the median number of gaseous satellites per cluster and consider the percent difference between a given cluster and its similar-massed companions. At all redshifts considered (here only showing redshifts $z\sim0,\,2$ for clarity), the amount of cool ICM (top left) and the cool ICM to total halo mass fraction (top right) increase with the number of gaseous satellites, at a fixed halo mass. That is, for a given cluster, the more gaseous satellites it hosts, the more cool gas it tends to have in its ICM. 

While at fixed redshift the cool ICM mass increases with the relative number of gaseous satellites, we want to understand how satellites can cause the cool ICM mass to decrease with time. Stripping of cool satellite gas via ram pressure becomes more effective at higher halo masses, which also increases with time. This is also reflected in, for example, the fraction of gaseous satellites that are jellyfish, which increases with cosmic time \citep{Zinger2024}. However, the total cool gas deposited into the ICM from satellites depends not only on the effective strength of cool gas stripping but also on the number of gaseous satellites themselves, which are known to undergo environmental effects and more so the longer they orbit in massive systems \citep[e.g.,][within the TNG model]{Donnari2021b}. We hence compute the number of satellites per cluster over the past $\approx13$~billion years (Fig.~\ref{fig:ICMCGM_Nsatellites} bottom panel). We plot the evolution for each of the 352 clusters from TNG-Cluster (thin curves) colored by their $z=0$ halo mass, and we include the medians within a $z=0$ halo mass bin (thick curves). At all times, the number of gaseous satellites increases with halo mass, a natural consequence of hierarchical structure formation. In contrast, the number of gaseous satellites per cluster decreases for all average cluster masses since redshifts $\sim1-2$. That is, while a cluster of mass $\sim10^{15}\, \msun$ today may host 10 gaseous satellites of stellar mass $>10^{9}\, \msun$, its progenitor at $z\sim1$ may have hosted twice as many. So while satellites are important sources of cool ICM \citep{Rohr2023}, they may become less crucial at later times due to their decreasing abundance.

We note that these results depend, in part, on the type of satellites that are being considered. In Appendix~\ref{app:satellites} we include two additional versions of Fig.~\ref{fig:ICMCGM_Nsatellites} using different definitions for satellites, and we summarize the results here. Recently, \citet{Chaturvedi2024} suggest that it is the number of massive satellites, not the total number of satellites, that is important when considering satellites as sources of cool ICM. When considering only gaseous satellites of stellar mass $>10^{10}\, \msun$ (instead of $>10^{9}\, \msun$), the same qualitative trends hold, and the average number of gaseous satellites per cluster decreases for all halo masses since $z\lesssim 1-2$. For this regime then, we conclude that satellite stellar mass may not be significant for the total amount of cool ICM, although it may still be important for the survivability of individual cool gas clouds \citep{Gronke2022,Roy2024}. We also consider removing the gaseous criterion -- that is, only requiring a stellar mass $>10^9\, \msun$ and cluster-centric distance $<\rvir$ -- and the stellar mass criterion -- that is, only requiring the subhalos to be within $\rvir$ (not shown). In TNG-Cluster, most $z=0$ cluster satellites of stellar mass $\sim10^{9-10}\, \msun$ are gas-poor and quenched, and those that are gas-rich tend to be recent infallers \citep{Rohr2024}. Thereby, the number of gaseous satellites proxies the number of recent infallers. However, the number of satellites (or subhalos in total) traces the overall hierarchical assembly of clusters and tends to increase monotonically with time, modulo satellites that have been tidally disrupted and satellite-satellite mergers. In these cases, the relative number of satellites still correlates with the cool ICM mass at a fixed cluster mass, but the number of satellites per cluster continues to increase with time until today. Therefore, while dark subhalos and gas-poor satellites are still important for the amount of cool ICM, where they may trigger gas cooling via density perturbations, the evolution of the total number of subhalos or satellites (gas-poor and gas-rich) alone cannot provide a potential explanation for why the cool ICM decreases towards $z=0$. We therefore conclude that at a fixed cluster mass and redshift, the cool ICM mass correlates with the number of gaseous satellites, and the decrease in the number of gaseous satellites since $z\sim1-2$ can possibly, at least partially, explain the decrease in the cool ICM.

\subsection{How SMBH feedback decreases the cool cluster gas mass} \label{sec:why_SMBH}

\begin{figure*}
    \includegraphics[width=\textwidth]{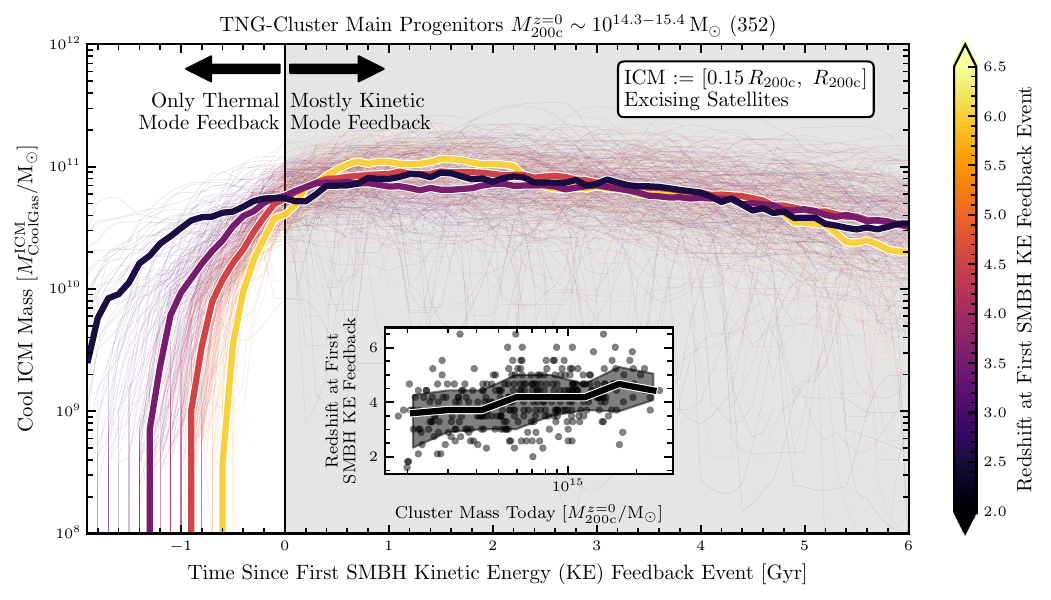}
    \includegraphics[width=\textwidth]{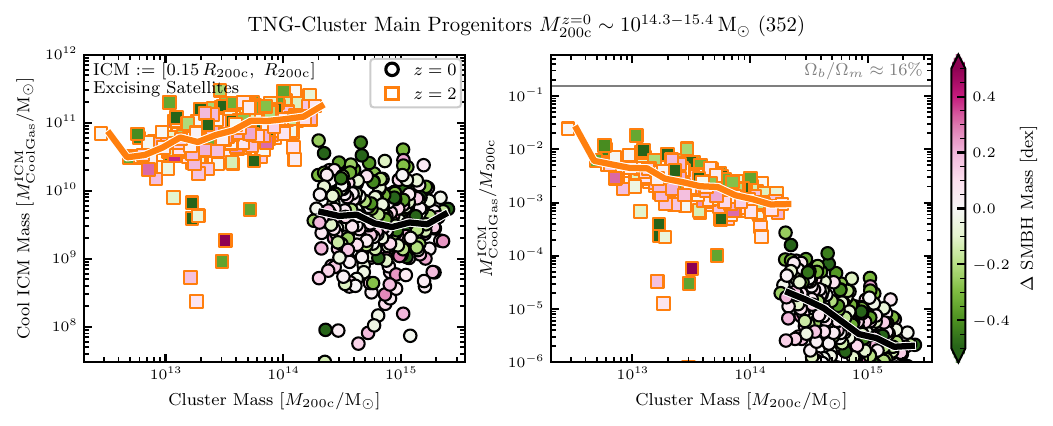}
    \caption{
    \textbf{Kinetic mode feedback from the SMBH causes the growth of the cool ICM mass to flatten and its total amount to decrease with time in TNG-Cluster.}
    \textit{Main Panel}: We show the evolution of the cool ICM mass since the first kinetic mode feedback event for all 352 clusters (thin curves), colored by the redshift at the first kinetic mode feedback event (medians as thick curves). Before this first kinetic mode event, all SMBH feedback was only in thermal mode, whereas afterwards a majority is in kinetic mode. In the inset we show that SMBHs in more massive clusters today tended to switch to kinetic mode at higher redshifts (earlier in cosmic time, although the effect is small at $\approx500$~Myr over one~dex in cluster mass today). Before the onset of kinetic mode feedback, the cool ICM mass increases with time; afterwards, the cool ICM mass tends to flatten and decrease until today. 
    \textit{Bottom Panels}: the effect of relative SMBH mass on the cool ICM mass and cool ICM to total halo mass fraction. Similar to Fig.~\ref{fig:ICMCGM_Nsatellites}, we now show how the relative difference in the central SMBH mass correlates with the cool ICM mass. At both redshifts considered here ($z\sim 2$, orange-outlined squares; $z\sim0$, black-outlined circles), clusters with undermassive SMBHs tend to have more cool ICM at fixed cluster mass. This effect is stronger at later times and demonstrates that feedback from SMBH affects the amount of cool gas in the ICM, at least according to TNG-Cluster.
    }
    \label{fig:first_smbh_rm}
\end{figure*}

\begin{figure}
    \includegraphics[width=\columnwidth]{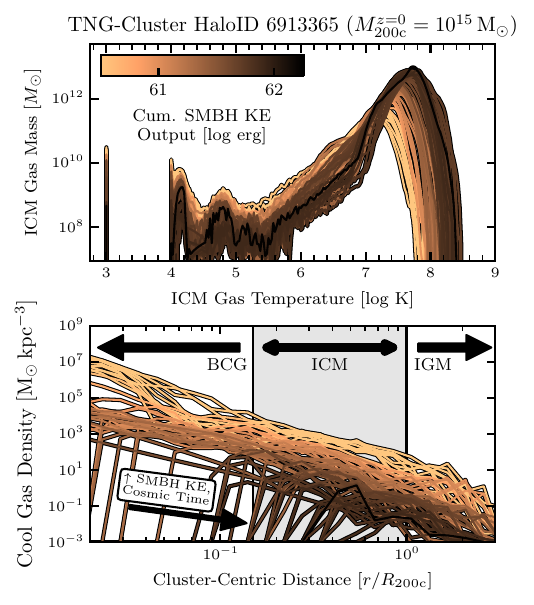}
    \caption{
    \textbf{Kinetic mode feedback from the central SMBH decreases the cool ICM mass}.
    Similar to Fig.~\ref{fig:example_evolution}, we now use the cumulative kinetic energy (KE) injected since birth from the central SMBH in kinetic mode as the color. As the cumulative energy output increases with time, the cool gas density at all radii decreases, including in the ICM, and the decrease occurs inside-out, suggesting that feedback from the SMBH drives the cool ICM mass. 
    }
    \label{fig:example_evolution_SMBH}
\end{figure}

Feedback from the central SMBH affects the ICM even out to, and at times extending beyond, the virial radius. In the TNG galaxy formation model, the kinetic, low-accretion mode feedback from SMBHs is largely responsible for quenching central galaxies \citep[e.g.,][]{Weinberger2018, Nelson2018} by both ejecting the interstellar medium gas and by offsetting the cooling times of the gaseous reservoirs in the halos \citep[e.g.,][]{Nelson2018b, Nelson2019,Truong2020,Davies2020, Zinger2020}. We henceforth examine exactly how the cool ICM mass changes once the central SMBHs starts to provide kinetic feedback. 

In Fig.~\ref{fig:first_smbh_rm}, for all 352 clusters (thin curves), we find the redshift when the main SMBH underwent a kinetic mode feedback event for the first time and plot the evolution of the cool ICM mass normalized to this time (main panel), colored by the redshift at the first kinetic mode feedback event. Before the onset of kinetic mode feedback, all clusters grew rapidly in their cool ICM mass. Interestingly at the onset of kinetic mode feedback, all clusters here have on average $\sim10^{10.5}\, \msun$ of cool ICM, regardless of when in cosmic time they in fact switch from thermal to kinetic mode feedback. This suggests that there may be a characteristic cool ICM mass at which the SMBHs tend to switch to kinetic mode feedback, even though these two quantities are spatially disconnected. In the first $\sim1-2$~Gyr after the onset of kinetic mode feedback, the growth of the cool ICM mass flattens out at approximately the same maximum cool ICM mass of $\sim10^{10.5-11}\, \msun$ regardless of when in cosmic time this occurs (the median trends are within the scatter of each other). 

We note that SMBHs in more massive $z=0$ clusters tended to have their first kinetic mode feedback event at higher redshfits, at earlier cosmic times (inset), although the effect is relatively small at $\approx500$~Myr difference across $\sim$~one dex in cluster mass. The central SMBHs of TNG-Cluster systems switched from thermal to kinetic mode in the redshift range $z=3-5$. They do so upon crossing some approximate halo mass threshold: as more massive $z=0$ clusters tend to cross this threshold at earlier times, their SMBHs also begin their kinetic mode feedback earlier (see also Fig.~\ref{fig:ICMCGM_MPBs}). In general after the first kinetic mode feedback event, SMBHs may switch between kinetic and thermal mode feedback; however, a majority of both the time and energy output in these massive systems occurs in kinetic mode and the majority are in kinetic mode feedback at $z=0$. Within the TNG model, we note that kinetic mode feedback, rather than the thermal mode feedback, is mainly responsible for affecting the low redshift ICM. 

Similarly to the number of satellites per cluster (Fig.~\ref{fig:ICMCGM_Nsatellites}), the average SMBH mass increases with cluster mass. Because of this and because the mass of a SMBH is proportional to the cumulative energy injected since birth via SMBH feedback, we compute the relative SMBH mass as the difference (in dex) between a given SMBH and the average SMBH mass for all clusters of a similar mass and correlate it with the cool ICM mass (Fig.~\ref{fig:first_smbh_rm} bottom panels). At a fixed cluster mass and redshift, clusters with undermassive SMBHs tend to have more cool ICM. The effect appears slightly stronger at later times. In Appendix~\ref{app:SMBH}, we repeat these panels colored instead by the cumulative kinetic energy output since birth and since the last snapshot ($\approx 150$~Myr), where the cumulative kinetic energy is stored in the simulation outputs. Briefly, the same qualitative trend holds for both of these SMBH properties, although there is more scatter, especially for the kinetic energy output since the last snapshot. We conclude that the cumulative kinetic mode feedback is more strongly correlated with the formation and/or survival of the cool ICM than recent feedback history.

Finally, we demonstrate in Fig.~\ref{fig:example_evolution_SMBH} how the ICM temperatures and cool gas radial profile evolve with the cumulative kinetic mode feedback since birth. Similar to the evolution with cosmic time (Fig.~\ref{fig:example_evolution}), the hot ICM mass increases and cool ICM mass decreases with cumulative kinetic mode feedback. This heating of the ICM is also related to the lengthening of the cooling times, functioning as a form of preventative feedback \citep[][]{Zinger2020,Voit2024}. We note that, in lower mass hosts, SMBH feedback tends to heat and redistribute the halo gas in the TNG model \citep[][]{Nelson2019,Zinger2020,Truong2021}. For these clusters, we then conclude that the SMBH feedback heats up the cool ICM inside-out and redistributes some of the ICM to larger distances, which potentially quenches star-formation \citep{Nelson2024}, drives X-ray cavities \citep[][]{Truong2024,Prunier2024}, and lengthens the cooling times of the hot ICM.

\section{Observational signatures of the cool ICM} \label{sec:disc}

\subsection{In-situ star formation in the ICM and \texorpdfstring{H$\alpha$}{Ha} emission} \label{sec:disc_sf}

\begin{figure*}
    \includegraphics[width=\textwidth]{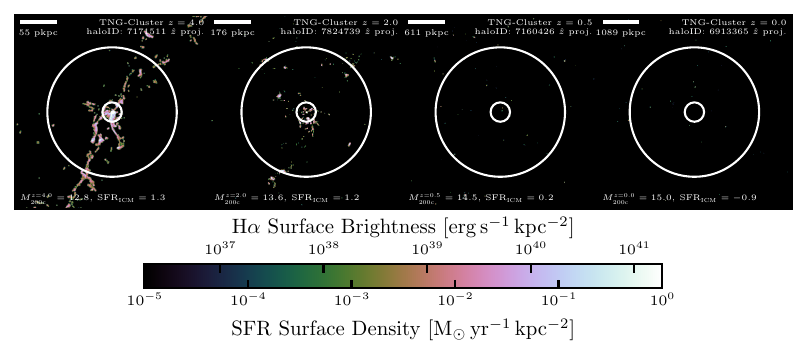}
    \includegraphics[]{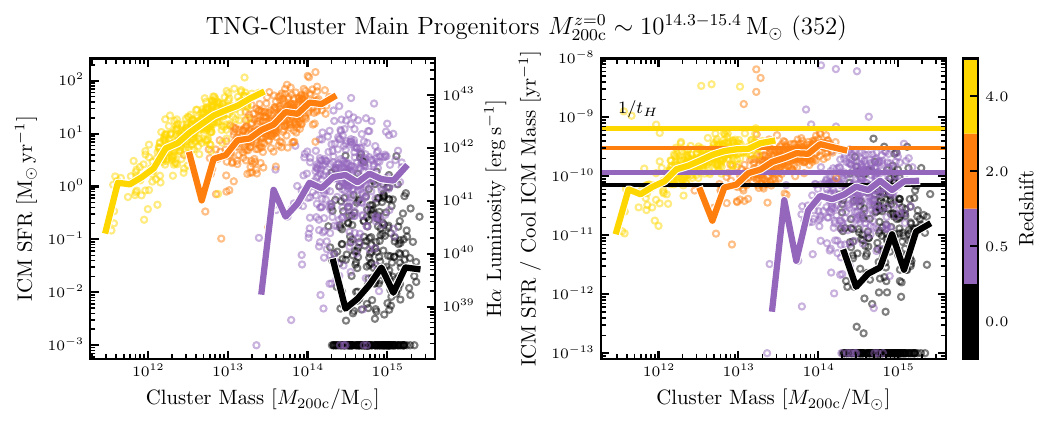}
    \caption{
    \textbf{Cluster progenitors in TNG-Cluster had (higher) in-situ ICM star formation rates than their descendants.}
    We include the H$\alpha$ surface brightness (top) and luminosity (bottom) using the unobscured SFR to H$\alpha$ conversion from \citet{Hao2011,Murphy2011,Kennicutt2012}; see text for details.
    \textit{Top Panels}: We show the evolution of the star formation rate surface density for an example cluster of $z=0$ mass $\sim 10^{15}\, \msun$, including all gas (including satellites) within a cube of $3\rvir$ centered on the BCG, where we project the gas cells using a cubic spline of variable kernel size according to the gas cell size. The annotations are as in Fig.~\ref{fig:CoolGasSurfaceDensity_mosaic}, and we additionally include the total ICM star formation rate in units of $[\log_{10} \msun\, {\rm yr}^{-1}]$ in the lower-left.
    \textit{Bottom panels}: The ICM SFR (left) and ICM SFR to cool ICM mass (right) as functions of cluster mass for all 352 clusters (circles), colored by redshift, where we include the median trend with mass as thick curves. We place clusters with ICM SFRs and ICM SFR to cool ICM mass below our resolution level at $10^{-3}\, \msun\, {\rm yr^{-1}},\ 10^{-13}\, {\rm yr^{-1}}$ respectively. We mark the the age of the universe $1/t_{\rm H}$ at each redshift as horizontal lines. The most extreme clusters at $z=2$ have ICM SFRs of $\sim10^{2}\, \msun\, \mathrm{yr^{-1}}$, corresponding to an H$\alpha$ surface brightness of $\sim10^{-17}\, \mathrm{erg\, s^{-1}\, cm^{-2}\, arcsec^{-2}}$, making it detectable with the Euclid Space Telescope or James Webb Space Telescope, according to TNG-Cluster.
    \label{fig:icmsfr}
    }
\end{figure*}

In-situ star formation in the ICM -- that is, star formation that occurs in ICM gas that is not bound to any satellite -- has been seen in early zoom-in simulations \citep{Puchwein2010,Mandelker2018}, in the most massive groups or low mass clusters of TNG50 \citep{Ahvazi2024b}, and in observations of runaway cooling clusters \citep{McNamara1989,Webb2015,Hlavacek-Larrondo2020}. The fate of this in-situ star formation is likely to become part of the intracluster light (ICL), perhaps by forming star clusters \citep{Mandelker2018,Ahvazi2024a}. However, to what degree such in-situ ICM star formation may contribute to the ICL remains observationally unconstrained and theoretically difficult to assess: the aforementioned studies suggest around $\approx10-30$~per~cent of the total ICL today \citep{Puchwein2010,Ahvazi2024b}, but we anticipate that this in fact may strongly depend on halo mass and other factors. In the following, we examine this phenomenon in TNG-Cluster.

We have seen that a non negligible amount of ICM gas in TNG-Cluster systems is star-forming, and that the star-forming ICM mass generally increases with redshift (Fig.~\ref{fig:ICMGasEvolution}, main panel). We now quantify further this in-situ ICM star formation. As a reminder, star formation in TNG-Cluster occurs in gas with densities $>0.1\, {\rm cm^{-3}}$ (see \S~\ref{sec:meth_tng}), with no additional physically motivated considerations for star formation to occur. We hence acknowledge from the onset that the following results likely depend on this simplified criterion, and that the ICM star formation rates may in fact change with different star formation models \citep[see also discussions in ][]{Puchwein2010,Ahvazi2024b}. 

With this caveat in mind, we examine a possible observable consequence: H$\alpha$ emission from star forming gas. Throughout, we convert the SFRs predicted in the TNG-Cluster simulated systems to H$\alpha$ luminosity via
\begin{equation}
    \log_{10} [{\rm H}\alpha / {\rm erg\, s^{-1}}] = \log_{10} [{\rm SFR} / \msun\, {\rm yr^{-1}}] + \log_{10}C_{{\rm H}\alpha}
\end{equation}
with a calibration factor $\log_{10}C_{{\rm H}\alpha} = 41.27$ \citep{Hao2011,Murphy2011,Kennicutt2012}.
We note that this simple approximation assumes that H$\alpha$ emission originates exclusively from star formation, that all star formation is totally unobscured (dust free), and that this conversion calibrated using star formation in local galaxies holds at high redshift and in the ICM. 

\subsubsection{Spatial distribution of the \texorpdfstring{H$\alpha$}{Ha} emission}

In the top panels of Fig.~\ref{fig:icmsfr}, we show the evolution since $z\lesssim4$ of the SFR surface density and H$\alpha$ surface brightness of an individual cluster from TNG-Cluster of $z=0$ mass $\sim10^{15}\, \msun$. In the images, we include all gas within a cube of size $3\rvir$, including satellites, centered on the BCG. Annotations are as in previous maps. In the bottom left of each stamp we include the halo mass $\mvir$ and the in-situ ICM SFR, which excludes gas bound to satellites, in units of $[\log_{10} \msun]$ and $[\log_{10} \msun\, {\rm yr^{-1}}]$, respectively. 

At $z=4$ (left panel), there is cluster wide star formation from within the BCG, throughout the ICM, and even stretching into the IGM. Inside the BCG, the star formation qualitatively appears to occur in a disk like structure. In the ICM, the star formation occurs in elongated structures in the directions of large scale filaments and the tails of stripped satellite galaxies. At $z=2$, there is still extended SFR in the ICM, although some of the star forming structures appear to be near satellite galaxies. At lower redshifts $z\lesssim0.5$, there is almost no extended star formation in the ICM, and nearly all star formation that does exist occurs in the interstellar media of satellite galaxies. However, at least some of the stripped intersteller media from satellite galaxies is expected to form stars in the TNG and TNG-Cluster simulations \citep{Goeller2023,Lora2024}. In fact, the total in-situ SFR in the ICM at $z=0$ is $\sim10^{-1}\, \msun\, {\rm yr^{-1}}$, and it is plausible to speculate that this gas stems from stripped satellites. 

The simulated cluster of Fig.~\ref{fig:icmsfr} is representative of the whole TNG-Cluster sample and, upon visual inspection of the maps, we believe that much of the star-forming ICM originates from satellite galaxies, which tend to be found co-spatially with filaments, where the star formation occurs at least partially in the elongated, ram pressure stripped tails of the satellites. This is consistent with the previous findings by \cite{Puchwein2010} but not with the claims of \cite{Ahvazi2024b}, which are based on the same galaxy formation model as our simulations. 
In particular, previously, \citet{Ahvazi2024b} find that, in the progenitors of the three most massive $z=0$ halos of TNG50 (total halo mass $\sim10^{13.7-14.3}\, \msun$), widespread ICM star formation occurs in small cloudlets within filamentary structures following the distribution of neutral hydrogen and loosely that of the underlying dark matter distribution. They suggest that the star-forming gas is not related to the ram pressure tails of stripped satellites. \citet{Puchwein2010}, on the other hand, suggest that a majority of the star-forming ICM in their cluster simulations of present day mass $\sim10^{14}\, \msun$ has distinct properties and origins compared to the rest of the ICM, where most of the star-forming ICM came from stripped satellites. 
More detailed analysis and comparisons will be required to clarify the situation. 

\subsubsection{Expected total \texorpdfstring{H$\alpha$}{Ha} luminosities}
In the bottom panels of Fig.~\ref{fig:icmsfr}, we extend the analysis of the in-situ ICM SFR to the total amount in all TNG-Cluster systems and their progenitors since $z\lesssim4$. We plot the in-situ ICM SFR (excluding satellites; left) and the ICM SFR to cool ICM mass ratio (right) as functions of cluster mass for all 352 clusters (circles), colored by redshift. We include the median mass trend at a fixed redshift as thick curves, and in the right panel we include the inverse age of the universe at that redshift as horizontal lines. We place clusters with SFRs below our resolution limit manually at $10^{-3}\, \msun\, {\rm yr^{-1}}$ (left) and $10^{-13}\, {\rm yr^{-1}}$ (right). 

At $z=0$, the ICM SFR has an approximately flat trend with cluster mass at $\sim10^{-2}\, \msun\, {\rm yr^{-1}}\ ({\rm H}\alpha \sim10^{39.5}\, {\rm erg\, s^{-1}})$. This median trend represents two populations of clusters today:
$\approx40$~per~cent of clusters have ICM SFRs $\lesssim10^{-3}\, \msun\, {\rm yr^{-1}}\ (\lesssim10^{38}\, {\rm erg\, s^{-1}})$; the other $\approx60$~per~cent of clusters have SFRs ranging up to $\sim1\, \msun\, {\rm yr^{-1}}\ (\sim10^{41}\, {\rm erg\, s^{-1}})$. For comparison, in the three most massive objects in TNG50, \citet{Ahvazi2024b} measure ICM SFRs of $\sim1\, \msun\, {\rm yr^{-1}}$, approximately an order of magnitude higher than the median ICM in TNG-Cluster, although still within the range of $z=0$ clusters. We note the objects studied by \citet{Ahvazi2024b} are less massive $\sim10^{13.7-14.3}\, \msun$ than the clusters in TNG-Cluster $\sim10^{14.3-15.4}\, \msun$. Additionally, the difference in resolution may play a role in the creation, survival, and potential star formation of cool clouds in the halo \citep{Nelson2020}. Namely, the mass resolution in TNG50 $\approx8.5\times10^4\, \msun$ is $\approx130$~times better than in TNG-Cluster at $\approx1.1\times10^7\, \msun$, and in general with the TNG galaxy formation model, star formation rates slightly increase with increasing resolution \citep{Pillepich2018b}.

At higher redshifts $z\gtrsim0.5$, the ICM SFR increases with cluster mass, and at a fixed cluster mass, the ICM SFR increases with redshift. Compared to a similar mass cluster at $z=0$, a $z=0.5$ cluster may have $\approx10-30$~times more star formation in the ICM. The most actively star-forming ICM at redshift $z\approx0.5$ have SFRs up to $\approx10^{1.5}\, \msun\, {\rm yr^{-1}}$, corresponding to an H$\alpha$ flux of $\sim 10^{-18}\, {\rm erg\, s^{-1}\, cm^{-2}\, arcsec^{-2}}$. At $z\approx2$, the most extreme objects may be have ICM SFRs of $\sim10^{2}\, \msun\, {\rm yr^{-1}}\ (\sim3\times 10^{43}\, {\rm erg\, s^{-1}})$, which corresponds to $\sim10^{-17}\, {\rm erg\, s^{-1}\, cm^{-2}\, arcsec^{-2}}$ in H$\alpha$.

Importantly, the in-situ star formation predicted by TNG-Cluster in the ICM of $z\approx2$ proto-clusters may be observable in the near-infrared within current surface brightness limits of the Euclid Space Telescope and James Webb Space Telescope, providing a test of star formation and galaxy formation models.

The ICM SFR to cool ICM mass ratio (right panel), akin to the star formation efficiency in nearby galaxies, details how efficient the cool ICM is at forming stars. Across the redshifts and cluster masses considered, the ICM SFR to cool ICM mass ratio increases with mass at a fixed redshift and vice versa; that is, the star-forming efficiency in the ICM increases with cluster mass at a fixed redshift, and with redshift at a fixed cluster mass. At $z=0$, the median ICM SFR to cool ICM mass ratio is significantly smaller than the inverse age of the Universe, by $\approx1-2$~dex,  meaning that it would take many Hubble times to deplete the cool ICM gas at that constant SFR. For redshifts $z\gtrsim0.5$, the median trends approach the inverse Hubble time, especially at large cluster masses at each redshift. The median star-forming efficiency in the ICM is maximum at $z\approx2$, where $\approx65$~per~cent of clusters have ICM SFR to cool ICM mass ratios greater than the inverse Hubble time. At all redshifts, there are individual clusters with ICM SFR to cool ICM mass ratio much larger than the inverse Hubble time.

\subsection{\texorpdfstring{The cool cluster gas in \ion{Mg}{ii} absorption}{The cool cluster gas in MgII absorption}} \label{sec:disc_abs}

\begin{figure*}
    \includegraphics[width=0.9\textwidth]{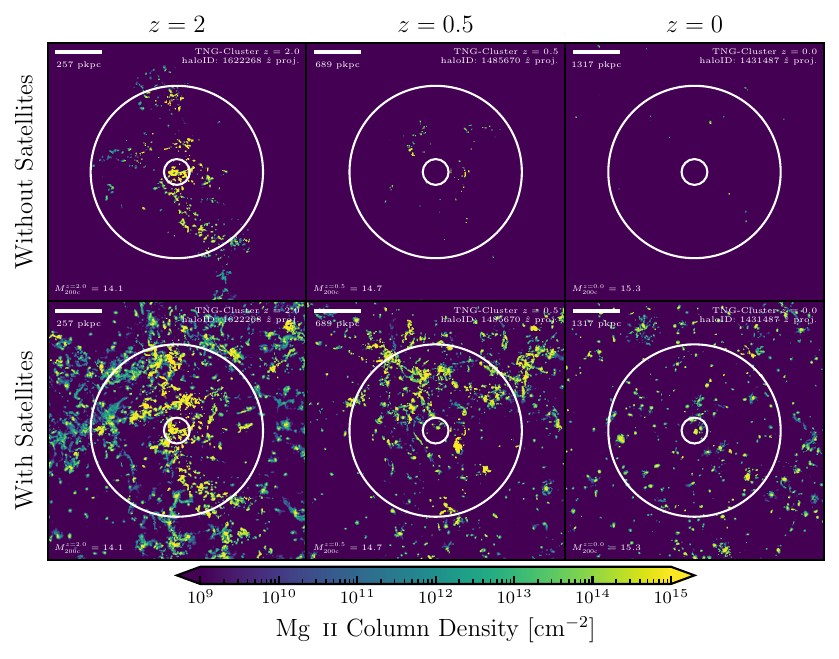}
    \includegraphics[width=\textwidth]{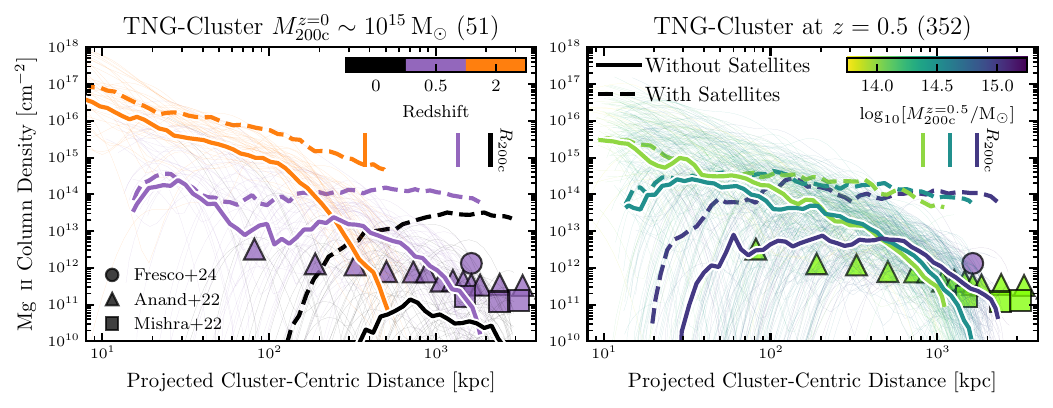}
    \caption{
    \textbf{Clusters and their progenitors contain \ion{Mg}{ii} gas in TNG-Cluster.}
    \textit{Top panels}: We show the evolution of the \ion{Mg}{ii} column density for an example cluster of $z=0$ mass $\sim 10^{15.3}\, \msun$, including gas (top: excluding satellites; bottom: including satellites and foreground+background halos) within a line-of-sight velocity $<2000\, {\rm km\, s^{-1}}$ of the BCG. The annotations and method are as in Fig.~\ref{fig:CoolGasSurfaceDensity_mosaic}.
    \textit{Bottom panels}: The \ion{Mg}{ii} column density radial profiles in clusters of present-day mass $\sim10^{15}\, \msun$ across cosmic time (left) and for all clusters of mass $\sim10^{13.75-15.25}\, \msun$ at $z=0.5$. The thin curves show the profiles for individual clusters, including gas within the a line of sight velocity $<2000\, {\rm km\, s^{-1}}$ of the BCG and excluding satellites, where we include the medians as thick curves. The dashed curves show the median profiles when also including satellites. We include stacked observational comparisons of clusters from \citet[][circle]{Fresco2024}, \citet[][triangles]{Anand2022}, and \citet[][squares]{Mishra2022}, which are colored to match the average redshift (left) and cluster mass (right); see text for details.
    }
    \label{fig:mgii_comparison}
\end{figure*}

Cool halo gas can also be detected as absorption features in spectra of background quasars or galaxies both in observations and simulations \citep{Peroux2020b}. Here we focus on \ion{Mg}{ii} in absorption to compare with recent observational studies with large sample statistics \citep{Mishra2022,Anand2022,Fresco2024}. 

We compute the number of \ion{Mg}{ii} atoms on a cell-by-cell basis, following \citet{Nelson2018,Nelson2024}. Briefly, we use the total magnesium mass as stored by the simulation and compute the ionization states using {\sc cloudy} \citep[][version c17]{Ferland2013,Ferland2017}, including both collisional and photoionization with the UV + X-ray background from \citet[][with the 2011 update]{Faucher-Giguere2009}, ensuring the self-consistency between the simulation and post-processing. We ignore local sources of radiation. {\sc cloudy} was run in single-zone mode and iterated to equilibrium, employing the fitting function for the frequency-dependent shielding from the background field at high density from \citet{Rahmati2013}. We then compute the column density of absorbers along a given line-of-sight (LoS) by projecting the clusters and summing the number of \ion{Mg}{ii} atoms. We include high resolution gas in the zoom simulation within a LoS velocity $<2000\, {\rm km\, s^{-1}}$ of the LoS velocity of the BCG, mimicking the selection from the observations (see below for details). In the TNG-Cluster simulation suite, however, the high-resolution zoom region does not sustain the path length corresponding to a Hubble expansion of $\pm2000\, {\rm km\, s^{-1}}$, which would be $>10\rvir$ from the BCG. We subsequently underestimate the total number of \ion{Mg}{ii} atoms originating from other satellites and halos, which may be significant in the projected column densities \citep{Rahmati2015,Nelson2018b,Weng2024}.

We present the results and comparisons with observations in Fig.~\ref{fig:mgii_comparison}. In particular, we provide quantification both with and without satellites and foreground and background objects. In the figure annotations by ``satellites'' we refer to both to gas bound to cluster satellites, as determined by the \subfind and Friends-of-Friends algorithms, and to gas from other halos that lie within the projected field of view and LoS velocity. That is, the ``with satellites'' results include both FoF satellites and foreground+background halos, commonly referred to as the ``2-halo'' or ``other-halo'' term \citep{Nelson2018b,Nelson2020}. Other studies may choose to separate these into different origins \citep[e.g.,][]{Weng2024}, or maybe even include other halos in fiducial choice \citep{Rahmati2015}. We exclude both gas bound to cluster satellites and to other halos, including the central and satellite galaxies of the other halos, because sightlines near satellites and other halos may be associated with these other objects rather than the cluster itself.

\subsubsection{Spatial distribution of \ion{Mg}{ii} absorbers}

In the top panels of Fig.~\ref{fig:mgii_comparison}, we show the evolution since $z=2$ of the \ion{Mg}{ii} column density for a single cluster of $z=0$ mass $\sim10^{15.3}\, \msun$, including all gas within a LoS velocity $<2000\, {\rm km\, s^{-1}}$ of the BCG, excluding satellites (top row) and including satellites (bottom row). For this example (and others as well, quantified below), satellites provide a large amount of \ion{Mg}{ii} absorbers in the ICM across redshifts and cluster centric distances, agreeing with expectations from lower mass hosts at $z=0$ \citep{Weng2024}. Clusters at lower redshifts $\lesssim0.5$ have few extended reservoirs of \ion{Mg}{ii} absorbers. Instead they tend to be small cloudlets throughout the ICM. When including satellites, the amount of \ion{Mg}{ii} absorbers increases, and there is some extended mission in the circumgalactic media or in the ram pressure stripped tails of these satellite galaxies. In general, the number of \ion{Mg}{ii} absorbers tends to increase with redshift. At $z\approx 2$, there are greater numbers of \ion{Mg}{ii} absorbers throughout the ICM and their covering fraction is higher. Without satellites, there exists a filamentary structure of \ion{Mg}{ii} absorbers, extending from within the BCG through the ICM to the IGM. When considering satellites, there are many such filaments connecting the satellites throughout the ICM. 

\subsubsection{\ion{Mg}{ii} column densities in the ICM}

In the bottom panels of Fig.~\ref{fig:mgii_comparison}, we compute the radial profiles of the \ion{Mg}{ii} column density for clusters across cosmic time (left) and mass (right). We include the radial profiles -- without satellites -- for individual clusters as thin curves colored by redshift (left) and mass (right) and include the median trend with redshift (left) and mass (right) as thick curves. We also include the median radial profiles when counting absorbers associated to satellites as dashed curves. We mark the average virial radius for each of the median radial profiles. Across all redshifts, cluster masses, and cluster-centric distances considered, satellites contribute to the total column density of \ion{Mg}{ii} absorbers, and their contribution increases with cosmic time, cluster mass, and cluster-centric distance. 

In the bottom left panel of Fig.~\ref{fig:mgii_comparison}, we examine the evolution of \ion{Mg}{ii} column density radial profiles for clusters in TNG-Cluster of present-day mass $\sim10^{15}\, \msun$ (51 clusters). A majority of present-day clusters (black) have little to no \ion{Mg}{ii} absorbers at small cluster-centric distance $\lesssim 100$~kpc, reflecting that many of these clusters have little to no cool gas in their centers \citep{Nelson2024,Lehle2024}. Their progenitors at $z\approx 0.5$ contained on average $\approx1-3$ orders of magnitude more \ion{Mg}{ii} absorbers throughout the ICM. At redshift $\approx 2$, there were large amounts of \ion{Mg}{ii} even in the cluster centers, reaching values of $\sim10^{16}\, {\rm cm}^{-2}$ at $\sim10$~kpc, reflecting that the average cluster progenitor at this time had cool gas in and around its core (Figs.~\ref{fig:coolgasprof_evolution},~\ref{fig:CoolGasSurfaceDensity_mosaic}). Here, there is a larger \ion{Mg}{ii} contribution from the ICM itself in the inner regions $\lesssim100$~kpc, and satellites mostly contribute at larger distances.The stacked observational data lie approximately between the median trends at redshifts 0 and 0.5 from TNG-Cluster out to $\approx \rvir$; we return to this comparison below. 

In the right panel, we show how the \ion{Mg}{ii} column density profile varies with cluster mass at a fixed time of $z\approx 0.5$. In the inner regions of clusters $\lesssim 100$~kpc, there is a clear separation between the low and high mass clusters in their numbers of absorbers. Namely, lower-mass clusters tend to have much higher column densities of \ion{Mg}{ii} absorbers $\sim10^{14-15}\, {\rm cm^{-2}}$ compared to high mass clusters $\sim10^{12-13}\, {\rm cm^{-2}}$, reflecting that the fraction of non-cool clusters in TNG-Cluster tends to increase with cluster mass \citep{Lehle2024}. In the outer regions $\gtrsim300$~kpc, there is little trend of the average column density of \ion{Mg}{ii} absorbers with mass until $\approx\rvir$ for the respective clusters, where the column density begins to drop exponentially. The observational data lie within the space covered by TNG-Cluster, although we caution a direct comparison.

\subsubsection{Preliminary comparisons to available observational results}

In the bottom panels of Fig.~\ref{fig:mgii_comparison}, we juxtapose our TNG-Cluster outcome to results from recent stacked observations of clusters from \citet[][circle]{Fresco2024}, \citet[][triangles]{Anand2022}, \citet[][squares]{Mishra2022}, which are colored to match the sample's average redshift (left) and cluster mass (right).

In general, the stacked observational data lie within the \ion{Mg}{ii} column density space probed by TNG-Cluster. From the simulations, we model the number of \ion{Mg}{ii} absorbers on a cell-by-cell basis for all clusters across distance, mass, and cosmic time. The observations rely on an equivalent width absorption feature on the background quasar spectrum through a given sightline near a cluster and then convert this equivalent width to a column density. Then by stacking tens to hundreds of thousands of these absorption features, either detected or non-detected, around different clusters at different projected distances, masses, and redshifts, one obtains the statistical properties of the cool ICM. 

Namely, \citet{Fresco2024} consider $\approx16,000$ quasar spectra from the (extended) Baryonic Oscillation Spectroscopic Surveys (BOSS \citealt{Dawson2013} and eBOSS \citealt{Dawson2016}) from the Sloan Digital Survey Survey (SDSS) data release 16 \citep[DR16][]{Lyke2020,Ahumada2020} around $\approx 1000$ X-ray selected clusters from the SPectroscopic IDentification of ERosita Sources \citep[SPIDERS;][]{Comparat2020,Clerc2020} at all projected distances $\lesssim 2$~Mpc, of mass $\sim10^{15}\, \msun$, and at redshifts $\sim0.3-0.6$ (average redshift $\approx0.41)$; \citet{Anand2022} cross-match $\approx 155,000$ known \ion{Mg}{ii} absorbers from SDSS DR 16 spectra \citep{Anand2021} with $\approx 70,000$ spectroscopically identified galaxy clusters from the Dark Energy Spectroscopic Instrument (DESI) Legacy Imaging Surveys \citep{Dey2019,Zou2021a} at projected cluster-centric distances $\lesssim 5$~Mpc, of mass $\sim$~a~few~$\times10^{14}\, \msun$, and at redshifts $\sim0.4-0.8$; \citet{Mishra2022} study $\approx81,000$ quasar-cluster pairs from $\approx64,000$ unique quasars from SDSS DR16 \citep{Lyke2020} around $\approx38,000$ clusters from the \citet{Wen2015} SDSS cluster catalog at projected distances $\sim1-4$~Mpc, of mass $\sim$~a~few$\times10^{14}\, \msun$, and at redshifts $\sim0.4-0.75$ (median redshift at $\approx0.55$). The studies from \citet{Fresco2024,Mishra2022} rely on stacking many spectra to obtain a statistical \ion{Mg}{ii} absorption feature, while \citet{Anand2022} utilize the $\approx 2700$ individually detected \ion{Mg}{ii} quasar-cluster absorbers identified by \citet{Anand2021}. The observed cluster samples are relatively complete, while TNG-Cluster is only volume-complete above cluster masses $>10^{15}\, \msun$ at $z=0$ (above masses $\gtrsim 10^{14.5}\, \msun$ at $z=0.5$; see \citealt{Nelson2024} for details).

The role of satellites, other halos, and intervening material at large distances are all critical for an apples-to-apples comparison with the observations. The observational studies tend to consider the \ion{Mg}{ii} absorbing gas within a LoS velocity of $\lesssim2000\, {\rm km\, s^{-1}}$ of the BCG, and we apply the same cut, additionally either removing satellites and other halos or including their contributions (but in no cases including the low resolution gas that would still be within a LOS velocity $\lesssim2000\, {\rm km\, s^{-1}}$ of the BCG). 

In cluster outskirts, TNG and TNG-Cluster satellites can both retain some of their own circumgalactic media \citep{Rohr2024} and begin depositing their metal-enriched ISM into the ICM \citep[e.g.,][]{Rohr2023,Zinger2024}, both of which may contribute to absorption features \citep{Weng2024}. \citet{Anand2022,Mishra2022} conclude that much of the cool, metal enrich gas in cluster outskirts is associated with satellites and may have been stripped from them in the past. This includes ISM, CGM, and (some) ram pressure stripped gas from the satellites. When we include gas associated with satellites and other halos, the average column density of \ion{Mg}{ii} absorbers in cluster outskirts $\gtrsim\rvir$ increases by orders of magnitude at all cluster masses and redshifts, confirming their significance to cool, metal-enriched cluster gas. ll, the observational data lie in the \ion{Mg}{ii} column density space predicted by the simulations, where the total \ion{Mg}{ii} column density (including satellites) from the simulations is higher than observed, but follow up studies that match the observational selection functions, forward-model the simulations to create mock spectra, and appropriately treat the contribution from cluster satellites and other halos are needed to fairly compare the observations with predictions from TNG-Cluster. Any discrepancies can shed light on the thermodynamic state of the ICM in the simulations and the associated role of SMBH feedback in forming and-or destroying \ion{Mg}{ii} in the cluster mass regime \citep[see also the discussion in][]{Nelson2024}.


\section{Summary and Conclusions} \label{sec:sum}

In this work, we analyze the intracluster medium (ICM) in 352 clusters of $z=0$ mass $\sim10^{14.3-15.4}\, \msun$ from the TNG-Cluster simulation suite across cosmic epochs. We focus on halo gas in the aperture $[0.15\rvir,\, \rvir]$, excluding satellites. We follow the progenitors of these clusters over the past $\sim13$~billion years, since $z\sim7$ and study the cool ICM of temperatures $\leq 10^{4.5}$~K, as opposed to the much more abundant and hot virial-temperature gas of $\sim10^{7-8}$~K. Such a study of the cool ICM over cosmic time and cluster mass for $\gtrsim 350$ clusters -- including processes and effects such as magnetic fields, gas cooling, SMBH feedback, satellite galaxies, and accretion from the large scale filaments -- is only possible with TNG-Cluster.

We summarize the main results and conclusions of our analysis:
\begin{itemize}
    \item According to TNG-Cluster, the cool ICM mass today is roughly constant across the considered cluster mass range. On the other hand, cluster progenitors unambiguously had more cool ICM, a higher cool ICM to total halo mass fraction, and a higher cool to total ICM mass fraction than their descendants today. At a fixed cluster mass, the cool ICM mass increases with increasing redshift, demonstrating that the cooler past of the ICM is due to more than just halo growth (\S~\ref{sec:results_coolpast}, Fig.~\ref{fig:ICMGasEvolution}).
    \item The cool cluster gas today is spatially scattered throughout the ICM in the form of more-or-less compact clouds, and a majority of the BCGs today have little to no cool gas. At higher redshifts ($z\gtrsim2$), large, cool filaments feed cool gas into the ICM from the IGM. Many of the higher-z BCGs are still star-forming and exhibit extended, cool gaseous disks. At all times, cool gas can be found both in and around satellites (\S~\ref{sec:results_spaceandtime}, Figs.~\ref{fig:coolgasprof_evolution},~\ref{fig:CoolGasSurfaceDensity_mosaic}).
    \item The azimuthally averaged cooling to free-fall time ratio in the ICM decreases with redshift and increases with cluster mass, as does the fraction of the hot ICM fulfilling the canonical $\tcool / \tff < 10$ criterion. This implies that the ICM in lower mass clusters and clusters at higher redshifts are more susceptible to cooling (\S~\ref{sec:why_cooling}, Fig.~\ref{fig:coolingtime}).
    \item The cool ICM mass correlates with the relative number of gaseous satellites at a fixed cluster mass and redshift, and this holds at all times since $z\lesssim 4$. The average number of gaseous satellites per cluster decreases by approximately a factor of two since redshifts $z\sim1-2$, potentially, but only partially, explaining why the cool ICM mass decreases with time (\S~\ref{sec:why_satellites}, Fig.~\ref{fig:ICMCGM_Nsatellites}).
    \item The onset and cumulative effect of SMBH kinetic mode feedback correlates with the time evolution of the cool ICM mass. The cool ICM mass is maximal at $\approx10^{10.5-11}\, \msun,\ \approx1-2$~Gyr after the onset of SMBH kinetic mode feedback, and then decreases to an average mass of $\sim10^{9.5}\, \msun$ today. Most SMBHs in TNG-Cluster tend to switch to kinetic mode feedback at a characteristic cool ICM mass of $\sim 10^{10.5}\, \msun$ at around $z\sim 3-5$, despite these two properties being spatially separated. At fixed redshift and cluster mass, the cool ICM mass increases in clusters hosting undermassive SMBHs. As the cumulative kinetic energy output from the SMBH increases, the cool ICM density decreases inside-out. We speculate that the cumulative kinetic energy from the SMBH, which has been shown within the TNG model to be the cause for star formation quenching in central and massive satellite galaxies, is likely the primary factor that sets the cool ICM mass (\S~\ref{sec:why_SMBH}, Figs.~\ref{fig:first_smbh_rm},~\ref{fig:example_evolution_SMBH}).
    \item Within the TNG-Cluster model, and quite directly depending on the choices of the star formation criteria therein, a non-negligible amount of ICM is star-forming, especially at higher redshifts. This star-forming gas potentially has observational signatures in the form of H$\alpha$ emission, where this in-situ ICM star formation reaches rates of $\sim10^2\, \msun\, {\rm yr^{-1}}$ at $z\approx2$, corresponding to a maximum H$\alpha$ surface brightness of $\sim10^{-17}\, {\rm erg\, s^{-1}\, cm^{-2}\, arcsec^{-2}}$, within current surface brightness limits of the Euclid Space Telescope and James Webb Space Telescope (\S~\ref{sec:disc_sf}, Fig.~\ref{fig:icmsfr}).
    \item The cool ICM of TNG-Cluster contains significant \ion{Mg}{ii} absorption, including large contributions from satellites and other halos. The column density of \ion{Mg}{ii} absorbers increases with redshift and decreases with cluster mass. Observations lie within the predicted range of TNG-Cluster, although more thorough comparisons are needed to draw stronger conclusions (\S~\ref{sec:disc_abs}, Fig.~\ref{fig:mgii_comparison}).
\end{itemize}

With this work, we demonstrate that current cosmological (magneto-)hydrodynamical simulations of galaxy formation and evolution like TNG-Cluster naturally return a multi-phase ICM, at least down to $\sim10^4$~K. In general, understanding how the cool ICM mass evolves is a multi-faceted, inter-connected problem, which we try to visualize in Fig.~\ref{fig:schematic} in \S~\ref{sec:why}. With this work, we extend previous findings on multi-phase gaseous halos in the TNG simulations, for the CGM of Milky Way-mass galaxies \citep{Ramesh2023b} and the intra-group medium of luminous red galaxies \citep{Nelson2020}, to the high-mass end. 

All the results and interpretations discussed in this paper are based on the TNG-Cluster simulation and underlying physical model, and hence there are a few caveats to consider. First, the effects of SMBH feedback on the halo gas may depend on the choice of feedback model: however, in general, many modern cosmological simulations implement some form of ejective and/or preventative SMBH feedback \citep[e.g.,][see also \citealt{Vogelsberger2020,Crain2023} for recent comparisons of hydrodynamic cosmological simulations of galaxies]{Davies2020,Terrazas2020,Ayromlou2023,Wright2024}. 
Additionally, some of the results related to the mass and size distribution of cool gas clouds, and hence total cool gas mass, are sensitive to numerical resolution. Specifically, the mass and covering fraction of the cool halo gas tend to increase, to varying degrees, with resolution even at TNG50 resolution, which is $\approx100$~times better than in TNG-Cluster \citep[e.g.,][]{Nelson2020}. Other simulations have suggested that higher resolutions enable smaller clouds to be resolved, potentially surviving longer, and that the total cool halo gas mass may increase with resolution, at least for Milky-Way-like halos \citep{Hummels2019,VandeVoort2019,Ramesh2024a}. These trends with resolution are true at redshifts $\gtrsim2$ as well \citep{Peeples2019,Suresh2019}, but it is unclear how, if at all, better resolution would affect the evolution of the cool ICM.
Lastly, there could be additional physical processes not modeled within the simulation, such as thermal conduction or cosmic rays, which may be important for the gas cooling and the survival of cool gas clouds. With these caveats in mind, we expect these results and interpretations to hold at least qualitatively across galaxy formation models, resolution levels, and in the real Universe, where future studies can be used to inform the models for the next generation of cosmological simulations.

Finally, a majority of our discussion assumes that there is an already existing cool gas reservoir at early cosmic epochs, which has been expected analytically and in early hydrodynamical simulations: for an unprecedented sample of massive clusters in a suite of cosmological galaxy-formation simulations, we show that these results still hold. Our analysis focuses on understanding how the cool ICM mass changes with time by studying the sources and sinks of cool gas and how their relative importance evolves. The TNG-Cluster simulation also yields, for example, observable predictions for the warm ($\sim10^{4.5-5.5}$~K) ICM related to emission or absorption of \ion{He}{ii}. To understand both the net effects on the cluster scale but also the small-scale effects on individual gas clouds within a cosmological context, we must combine efforts and results across scales of simulations, continuing to learn and inform models via comparisons with observations. Understanding the spatial and temperature structure of halo gas across redshifts is a viable next step to constraining our models of galaxy formation and evolution.

\section*{Acknowledgments}

ER is a fellow of the International Max Planck Research School for Astronomy and Cosmic Physics at the University of Heidelberg (IMPRS-HD). AP acknowledges funding from the European Union (ERC, COSMIC-KEY, 101087822, PI: Pillepich). DN and MA acknowledge funding from the Deutsche Forschungsgemeinschaft (DFG) through an Emmy Noether Research Group (grant number NE 2441/1-1). ER would like to thank colleagues Eduardo Ba{\~n}ados Urmila Chadayammuri, Seok-Jun Chang, Dimitris Chatzigiannakis, Callie Clontz, Max H{\"a}berle, Julie Hlavacek-Larrondo, Katrin Lehle, Marine Prunier, Nico Winkel, Simon White, and Zhang-Liang Xie for insightful conversations that improved the analysis. The authors would like to thank the anonymous referee for the thoughtful comments that strengthened and improved the clarity of the manuscript.

This research was supported by the International Space Science Institute (ISSI, \url{https://www.issibern.ch/}) in Bern, through ISSI International Team project \#564 (The Cosmic Baryon Cycle from Space).

The TNG-Cluster simulation suite has been executed on several machines: with compute time awarded under the TNG-Cluster project on the HoreKa supercomputer, funded by the Ministry of Science, Research and the Arts Baden-Württemberg and by the Federal Ministry of Education and Research; the bwForCluster Helix supercomputer, supported by the state of Baden-Württemberg through bwHPC and the German Research Foundation (DFG) through grant INST 35/1597-1 FUGG; the Vera cluster of the Max Planck Institute for Astronomy (MPIA), as well as the Cobra and Raven clusters, all three operated by the Max Planck Computational Data Facility (MPCDF); and the BinAC cluster, supported by the High Performance and Cloud Computing Group at the Zentrum für Datenverarbeitung of the University of Tübingen, the state of Baden-Württemberg through bwHPC and the German Research Foundation (DFG) through grant no INST 37/935-1 FUGG. 

\section*{Data Availability and Software Used}
The TNG-Cluster simulation will be made public one year after the first publications by the team, probably in the first quarter of 2025. The data will be organized as for all other TNG simulations, which are already publicly available at \url{https://www.tng-project.org/} and described in \citet{Nelson2019}. All codes used to analyze the TNG-Cluster data and to produce the figures in this paper are publicly available at \url{https://github.com/ecrohr/TNG\_RPS}. 

Software used: {\sc Python} \citep{VanDerWalt2011}; {\sc Numpy} \citep{VanDerWalt2011,Harris2020}; {\sc Scipy} \citep{Virtanen2020}; {\sc Matplotlib} \citep{Hunter2007}; {\sc Jupyter} \citep{Kluyver2016}.

This work made extensive use of the NASA Astrophysics Data System and \url{https://arxiv.org/} preprint server.

\bibliographystyle{mnras}
\bibliography{references}


\appendix

\section{The effect of different normalizations on cool gas radial profiles} \label{app:normalizations}

\begin{figure*}
    \includegraphics[width=\textwidth]{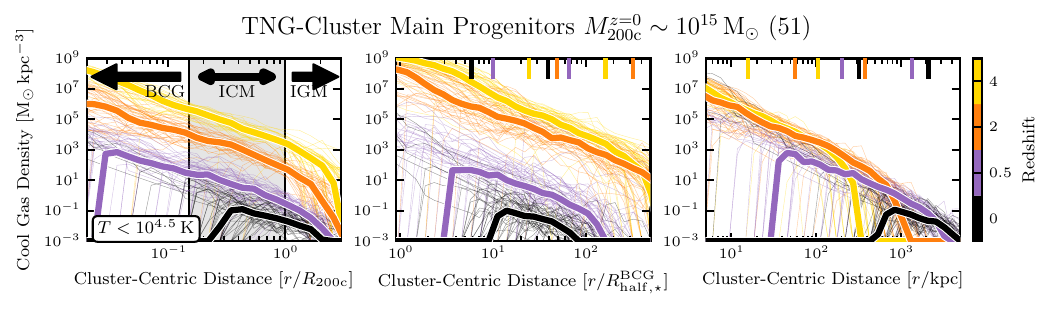}
    \caption{
    The effect of different normalizations on cool gas radial profiles. 
    From left to right, the radial profiles are normalized by the virial radius, normalized by the BCG stellar half mass radius, and in physical coordinates. For the latter two, we mark on the top $x$-axis where the average adopted BCG-ICM and ICM-IGM boundaries are at $0.15\rvir,\, \rvir$ respectively. All profiles are only for clusters whose $z=0$ mass is $\sim10^{15}\, \msun$ (51 clusters; thin curves), and we include the medians at each redshift as thick curves. At all overlapping radii and redshifts since $z\sim 4$, cluster progenitors had more cool gas than their descendants today.
    }
    \label{fig:normalizations}
\end{figure*}

In Fig.~\ref{fig:normalizations} we show how cool gas radial profiles for clusters of $z=0$ mass $\sim10^{15}\, \msun$ evolve with redshift, as was shown in Fig.~\ref{fig:coolgasprof_evolution}. Here, we show the effect of different normalizations on the cool gas radial profiles. From left to right, the profiles are normalized by the virial radius $\rvir$ (this figure is the same as Fig.~\ref{fig:coolgasprof_evolution}), by the stellar half mass radius of the BCG $\rhalfstar^{\rm BCG}$, and in physical coordinates. For the the latter two, we mark on the top $x$-axis where the positions of the average adopted BCG-ICM and ICM-IGM boundaries at $0.15\rvir, \rvir$, respectively. 

For all normalizations considered, the cool gas density decreases as an approximate power law in the ICM and exponentially in the IGM; lower-redshift clusters tend to have little cool gas in their BCGs while their higher-redshift progenitors tend to have cool gas in their cores.
When normalizing the radial profiles by the stellar half mass radius of the BCG (center panel), the cool gas density increase with redshift at all cluster-centric distance. However, the definition of the ICM in units of the stellar half mass radius changes significantly across redshifts, which likely reflects a morphological transformation from discy star-forming galaxies at higher redshift to an elliptical, quenched BCG today. In physical units (right panel), the cool gas density increases with redshift in the ICM regions, but there is little overlap between the redshifts in physical spaces due to the halos growing in size with time.

\section{The effect of cluster mass on the cool gas radial profiles today} \label{app:halo_mass}

\begin{figure}
    \includegraphics[width=\columnwidth]{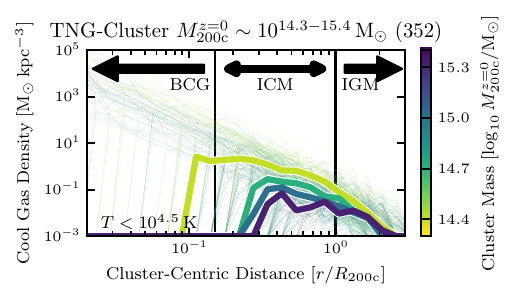}
    \includegraphics[width=\columnwidth]{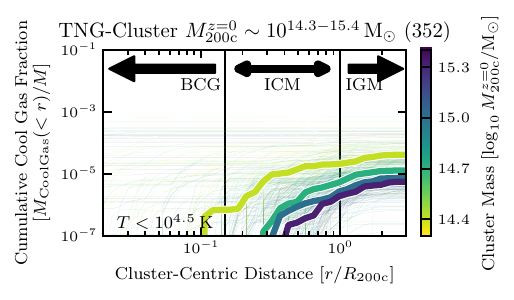}
    \caption{
    Similar to Fig.~\ref{fig:coolgasprof_evolution} but for all clusters at $z=0$. We plot the cool gas density (top) and cumulative cool gas mass fraction (bottom) radial profile, normalized by the virial radius, for all 352 clusters today (thin curves) colored by their mass, and we include the medians at a fixed mass (thick curves). Lower mass clusters tend to have higher cooler gas densities and cool gas fractions at all radii considered at $z=0$ in TNG-Cluster.
    }   
    \label{fig:coolgasradprof_z0}
\end{figure}

In Fig.~\ref{fig:coolgasradprof_z0} we show the effect of $z=0$ cluster mass on the cool gas radial profiles, in a similar fashion to Fig.~\ref{fig:coolgasprof_evolution}, for all 352 clusters in TNG-Cluster. We show the cool gas density (top) and the cumulative cool gas mass fraction (bottom) radial profiles, normalized by the virial radius. Lower-mass clusters tend to have higher cool gas densities in their ICM, although the effect is small compared to trend with redshift (Fig.~\ref{fig:coolgasprof_evolution}). There are a number of clusters with non-negligible cool gas masses in their BCGs, which are likely cool-core clusters and are potentially even star-forming \citep{Lehle2024,Nelson2024}.

\section{Different definitions of satellites and their correlations with cool ICM mass} \label{app:satellites}

\begin{figure*}
    \includegraphics[]{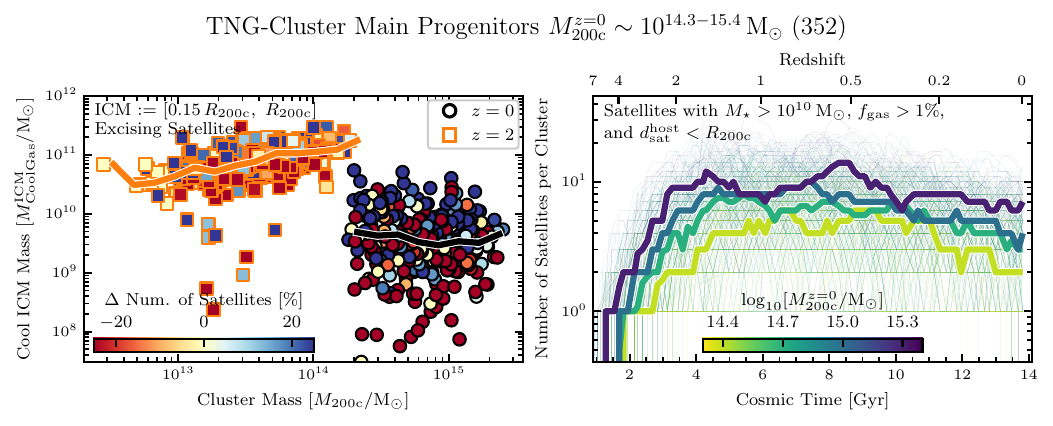}
    \includegraphics[]{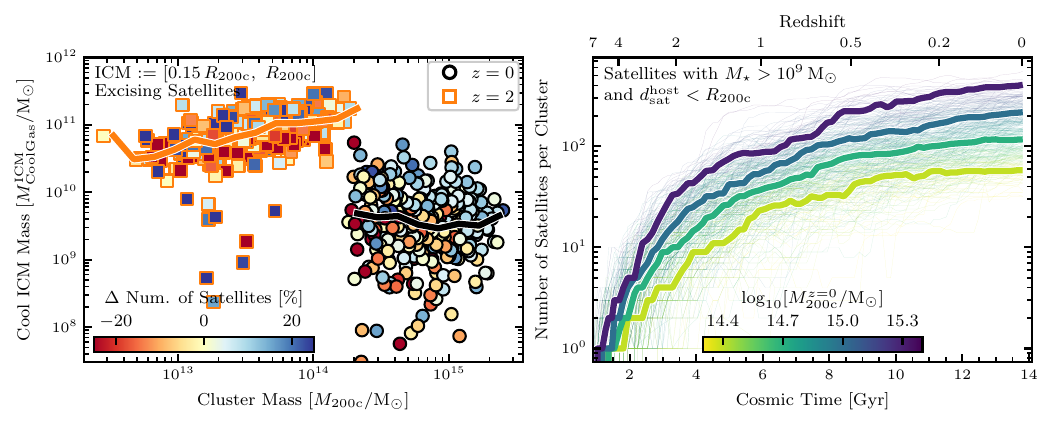}
    \caption{
    Similar to Fig.~\ref{fig:ICMCGM_Nsatellites}, now only considering massive satellites of stellar mass $>10^{10}\, \msun$ (top panels) or all satellites of stellar mass $>10^9\, \msun$ regardless of their gas content (that is, including gas-poor galaxies as well; bottom panels). 
    The results are qualitatively similar when only considering massive satellites, so we conclude that is not only massive satellites that are relevant to determining the cool ICM mass. The correlations of cool ICM mass with relative number of all satellites regardless of their gas content are similar (albeit slightly weaker) to the fiducial choice (only gaseous satellites), but the number of satellites monotonically increases with time, and thereby is not as helpful in explaining why the cool ICM mass decreases.
    }
    \label{fig:ICMCGM_Nsatellites_app}
\end{figure*}

In \S~\ref{sec:why_satellites} and Fig.~\ref{fig:ICMCGM_Nsatellites} we demonstrate that the cool ICM mass increases with the number of gaseous satellites -- namely satellites with stellar mass $>10^{9}\, \msun$ and gas to total baryonic mass fraction $f_{\rm gas} = \mgassat / \mstarsat > 1$~per~cent -- and relative to similar mass clusters at the same redshift. We also show that the average number of gaseous satellites per cluster decreases by factors of $\approx1.5-2$ since redshifts $\lesssim1-2$, partially explaining the cool ICM mass decreases with cosmic time. As a reminder, the majority of TNG-Cluster satellites are gas-poor at $z=0$, and more recent-infallers and more massive satellites are more likely to be gas-rich \citep{Rohr2024}. Therein, considering gaseous satellites is a proxy for a combination of recent-infalling and massive satellites, not necessarily the total number of satellites. Gaseous satellites are able to directly deposit their cool ISM into the ICM via ram pressure stripping and outflows, in addition to causing hydrodynamical instabilities that lead to gas cooling \citep[e.g.,][]{Rohr2023}. Recently, \citet{Chaturvedi2024} suggest that massive satellites, not all satellites, are significant sources of cool gas in the ICM, and that only massive satellites deposit large enough clouds of cool gas to survive for cosmological timescales in the ICM \citep{Gronke2022,Roy2024}. Moreover, gas-poor satellites, or even dark subhalos without baryonic material, may still be able to cause density perturbations leading to gaseous cooling. We now consider different definitions of satellites in Fig~\ref{fig:ICMCGM_Nsatellites_app}.

These figures are similar in style as Fig.~\ref{fig:ICMCGM_Nsatellites} (here only showing the correlations with the cool gas mass), where the left panels show how the number of satellites relative to similar mass halos affect the cool ICM mass as functions of cluster mass and redshift; the right panels shows the time evolution of the number of satellites per cluster over the past $\approx13$~billion years, since redshift $\lesssim7$. In the top panels of Fig.~\ref{fig:ICMCGM_Nsatellites_app} we consider massive, gaseous satellites with stellar mass $\mstarsat > 10^{10}\, \msun$ and gas to total baryonic mass fraction $>1$~per~cent. The cool ICM mass tends to increase with relative number of massive satellites, which is qualitatively consistent when considering gaseous satellites with stellar mass $>10^9\, \msun$. Additionally, the average number of massive, gaseous satellites decreases by factors of $\approx1.5-2$ since redshift $\lesssim1$, although there are low number statistics, again similarly to the number of gaseous satellites with stellar mass $>10^9\, \msun$. Due to the similarities in these results, we conclude that not only massive satellites are relevant to determining the cool ICM mass and its decline with cosmic time, but gaseous satellites all together. 

In the bottom panels of Fig.~\ref{fig:ICMCGM_Nsatellites_app} we consider all satellites with stellar mass $>10^9\, \msun$, regardless of their gaseous content. The results are qualitatively similar with both definitions of satellites. The cool ICM mass still tends to increase with the relative number of satellites, but the trends become qualitatively weaker than when considering gaseous satellites. Unlike when considering gaseous satellites, however, here the average number of satellites monotonically increases with cosmic time. In this sense, the evolution of the number of satellites does not reflect the evolution of the cool ICM mass, and thereby cannot help explain its decrease since redshifts $\approx2-3$. The same results qualitatively hold when considering all subhalos, regardless of their baryonic content (not shown). However, the total number of satellites, especially above a stellar mass or brightness threshold, follows the evolution of the total cluster mass and can be calibrated to estimate cluster mass, known as the mass-richness relation \citep[e.g.,][see also fig.~1 from \citealt{Rohr2024}]{Costanzi2019,Abdullah2020,Abdullah2023}.

\section{Different SMBH Properties and their effects on the cool ICM} \label{app:SMBH}

\begin{figure*}
    \includegraphics[]{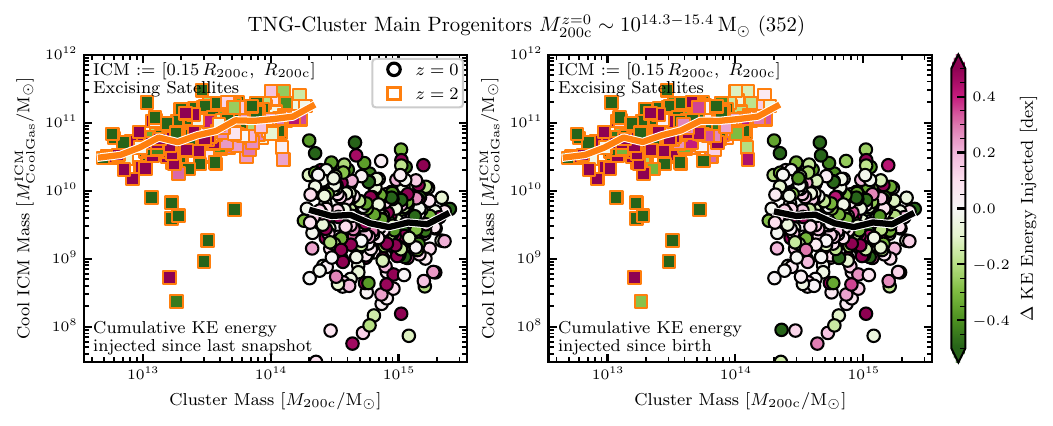}
    \caption{
    Similar to Fig.~\ref{fig:first_smbh_rm} bottom panels, but showing two additional SMBH properties: the amount of kinetic energy injected since the last snapshot (left) and the cumulative kinetic energy injected since birth (right). The qualitative results are similar, albeit stronger for the SMBH mass and cumulative energy since birth than for the energy injected since the last snapshot, suggesting that the cumulative SMBH kinetic feedback is more important than the recent feedback history in affecting the cool ICM mass. 
    }
    \label{fig:ICMCGM_SMBH_alt}
\end{figure*}

In \S~\ref{sec:why_SMBH} and Fig.~\ref{fig:first_smbh_rm} we demonstrate how the onset of kinetic energy feedback from the central SMBH coincides with the maximum cool ICM mass and that the cool ICM mass correlates with the SMBH mass relative to similar mass clusters. In Fig.~\ref{fig:ICMCGM_SMBH_alt} we consider how the cool ICM mass correlates with two alternative SMBH properties: the cumulative kinetic energy output since the previous snapshot (left) and the total cumulative kinetic energy output since birth (right). The kinetic energy output since the previous snapshot, which corresponds to $\approx150$~Myr, proxies recent SMBH activity and feedback, while the latter proxies the total cumulative output, similar to the SMBH mass. There exist weak trends with the kinetic energy output since the last snapshot, but the trends with the cumulative kinetic energy output and SMBH mass are more apparent. We therefore conclude that it is the cumulative kinetic energy from the SMBH, likely not the recent SMBH feedback, that sets the cool ICM mass, especially at later times.

\label{lastpage}
\end{document}